%% file: entint.tex
\documentclass[letterpaper,floatfix,reprint,aps,pre,superscriptaddress]{revtex4-1}

\usepackage{graphicx}
\usepackage{amssymb,amsmath,amsfonts}
\usepackage[T1]{fontenc}
\usepackage{times}
\usepackage{hyperref}
\usepackage{array}

\newcolumntype{M}{>{$\vcenter\bgroup\hbox\bgroup}c<{\egroup\egroup$}}
\newcolumntype{T}{>{$\vtop\bgroup\hbox\bgroup}c<{\egroup\egroup$}}

\usepackage{ifthen}

\usepackage{tikz}
\usetikzlibrary{arrows,shapes,positioning,calc,3d}
\usepackage{makecell}
\usepackage{youngtab}

\newcommand{\bq}{\mathbf{q}}

\newcommand{\eg}{\textit{e.g.}}
\newcommand{\Eg}{\textit{E.g.}}
\newcommand{\ie}{\textit{i.e.}}

\DeclareMathOperator{\Tr}{Tr}

\definecolor{gcb1}{HTML}{A6CEE3}
\definecolor{gcb2}{HTML}{1F78B4}
\definecolor{gcb3}{HTML}{B2DF8A}
\definecolor{gcb4}{HTML}{33A02C}
\definecolor{gcb5}{HTML}{FB9A99}
\definecolor{gcb6}{HTML}{E31A1C}
\definecolor{gcb7}{HTML}{6A3D9A}
\definecolor{gcb8}{HTML}{FF7F00}
\definecolor{gcb9}{HTML}{CAB2D6}
\definecolor{gcb10}{HTML}{6A3D9A}

\begin{document}
\preprint{arXiv:1309.1187}
\title{Understanding shape entropy through local dense packing}
\author{Greg \surname{van Anders}}
\affiliation{Department of Chemical Engineering,
University of Michigan, Ann Arbor, MI 48109-2136, USA}
\author{Daphne \surname{Klotsa}}
\affiliation{Department of Chemical Engineering,
University of Michigan, Ann Arbor, MI 48109-2136, USA}
\author{N.\ Khalid \surname{Ahmed}}
\affiliation{Department of Chemical Engineering,
University of Michigan, Ann Arbor, MI 48109-2136, USA}
\author{Michael \surname{Engel}}
\affiliation{Department of Chemical Engineering,
University of Michigan, Ann Arbor, MI 48109-2136, USA}
\author{Sharon C.\ \surname{Glotzer}}
\affiliation{Department of Chemical Engineering,
University of Michigan, Ann Arbor, MI 48109-2136, USA}
\affiliation{Department of Materials Science and Engineering,
University of Michigan, Ann Arbor, MI 48109-2136, USA}

\begin{abstract}
Entropy drives the phase behavior of colloids ranging
from dense suspensions of hard spheres or rods to dilute suspensions of hard
spheres and depletants. Entropic ordering of anisotropic shapes into complex
crystals, liquid crystals, and even quasicrystals has been demonstrated recently
in computer simulations and experiments. The ordering of shapes appears to arise
from the emergence of directional entropic forces (DEFs) that align neighboring
particles, but these forces have been neither rigorously defined nor quantified
in generic systems. Here, we show quantitatively that shape drives the phase
behavior of systems of anisotropic particles upon crowding through DEFs. We
define DEFs in generic systems, and compute them for several hard particle
systems. We show that they are on the order of a few $k_\mathrm{B}T$ at the
onset of ordering, placing DEFs on par with traditional depletion, van der
Waals, and other intrinsic interactions. In experimental systems with these
other interactions, we provide direct quantitative evidence that entropic
effects of shape also contribute to self-assembly. We use DEFs to draw a
distinction between self-assembly and packing behavior. We show that the
mechanism that generates directional entropic forces is the maximization of
entropy by optimizing local particle packing. We show that this mechanism occurs
in a wide class of systems, and we treat, in a unified way, the entropy-driven
phase behavior of arbitrary shapes incorporating the well-known works of
Kirkwood, Onsager, and Asakura and Oosawa.
\end{abstract}
\maketitle

\begin{center}
  \colorbox{lightgray}
  {
  \begin{minipage}{8.3cm}
    \textbf{Significance}\\

    Many natural systems are structured by the ordering of repeated, distinct
    shapes. Understanding how this happens is difficult because shape affects
    structure in two ways. One is how the shape of a cell or nanoparticle, for
    example, affects its surface, chemical, or other intrinsic properties. The
    other is an emergent, entropic effect that arises from the geometry of the
    shape itself, which we term ``shape entropy'', and is not well understood.
    In this paper, we determine how shape entropy affects structure. We
    quantify the mechanism, and determine when shape entropy competes with
    intrinsic shape effects. Our results show that in a wide class of systems
    shape affects bulk structure because crowded particles optimize their
    local packing.
  \end{minipage}
  }
\end{center}

\section{Introduction}
\begin{figure*}
  \begin{center}
  \includegraphics[width=17cm]{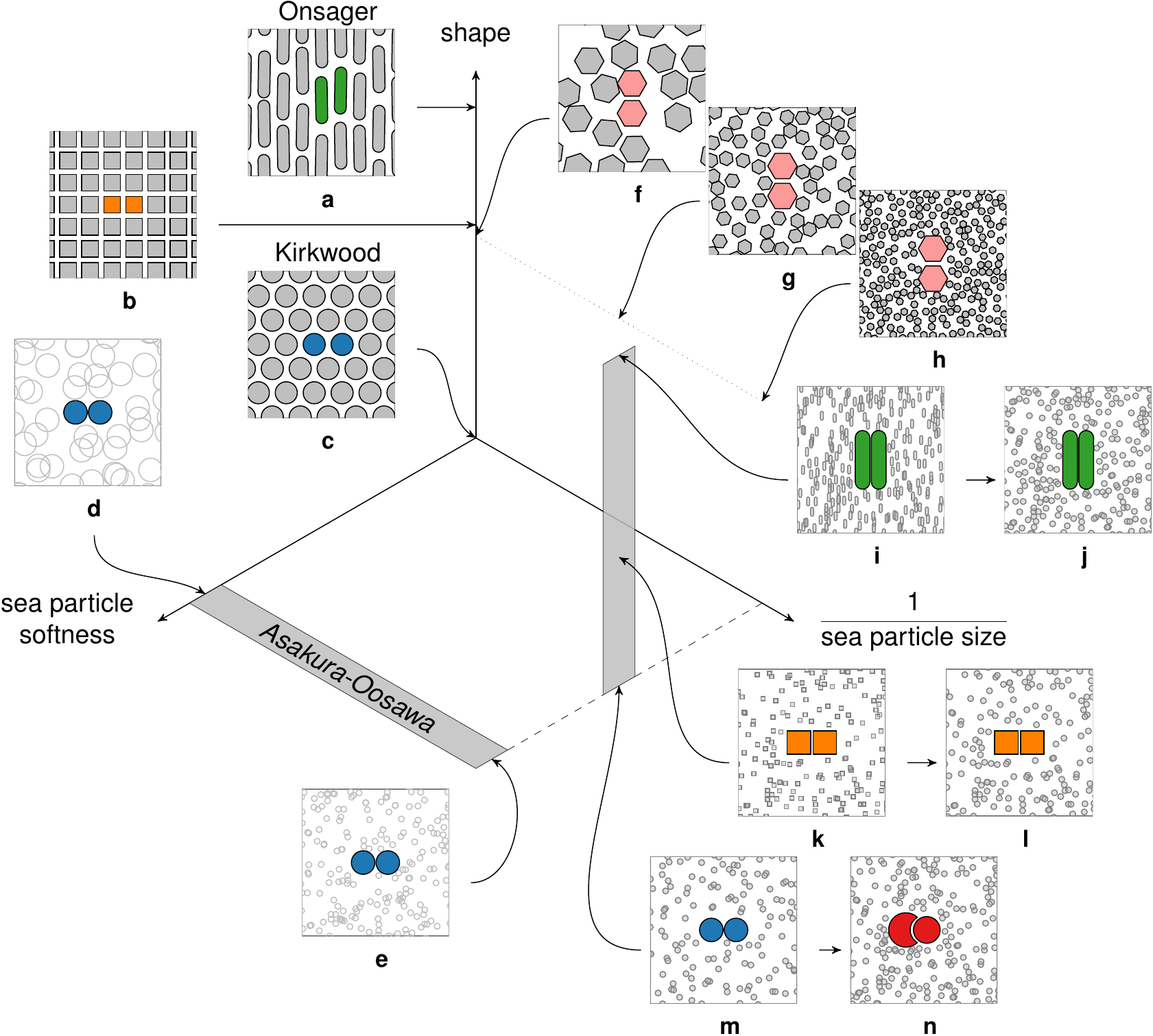}
  \end{center}
  \caption{The general nature of entropic interactions treated in this work
  applies to a broad class of known systems. Here we represent them on three
  orthogonal axes. One axis represents, schematically, the shape of the
  constituent particles, with spheres at the origin. The other two axes concern
  the sea of particles that are being integrated out and provide the effective
  interaction. On one axis is the inverse of the strength of the interaction
  between them (where 0 represents hard steric exclusion). On the other axis is
  the ratio of the characteristic size of the particles of interest to that of
  the particle being integrated out. Other axes, not shown, represent the shape
  of the particle being integrated out, mixtures of particle shapes and types,
  {\it etc.} Examples of known experimental and model systems are sketched to
  illustrate their location on these axes (see references and description in
  text). Different coloring schemes are used to indicate penetrable sea particles
  \protect\tikz \protect\draw[draw=lightgray,very thin,fill=white] (0.7ex,0.7ex)
  circle (0.7ex); ,
  semi-penetrable sea particles
  \protect\tikz \protect\draw[draw=gray,very thin,fill opacity=0.5,fill=lightgray]
  (0.7ex,0.7ex) circle (0.7ex); ,
  and impenetrable sea particles
  \protect\tikz \protect\draw[draw=black,very thin,fill=lightgray] (0.7ex,0.7ex)
  circle (0.7ex); .
  \label{fig-scope}
  }
\end{figure*}

Nature is replete with shapes. In biological systems,
eukaryotic cells often adopt particular shapes, for example, polyhedral
erythrocytes in blood clots \cite{polyhedrocytes}, and dendritic neurons
in the brain \cite{neuroscience}. Before the development of genetic techniques
prokaryotes were classified by shape, as bacteria of different shapes were
implicated in different diseases \cite{microbio}. Virus capsids
\cite{brookscapsid,monicabuckle} and the folded states of proteins
\cite{molbiocell4} also take on well-recognized, distinct shapes. In
non-living systems, recent advances in synthesis make possible granules,
colloids, and nanoparticles in nearly every imaginable shape
\cite{glotzsolomon,yangrev,steinrev2,shapecolloids,shape4sa,xiarev}.
Even particles of nontrivial topology are now possible \cite{topocolloid}.

The systematic study of families of idealized colloidal- and nanoscale systems
by computer simulation has produced overwhelming evidence that shape is
implicated in the self-assembly \footnote{We use self-assembly to apply to
(thermodynamically) stable or metastable phases that arise from systems
maximizing their entropy in the presence of energetic and volumetric constraints
(temperature and pressure, \ie\ spontaneous self-assembly) or other constraints
(\eg\ electromagnetic fields, \ie\ directed self-assembly).} of model systems of
particles.
\cite{dijkstranonconvex,trunctet,dijkstrasuperballs,zoopaper} In these model
systems, the only intrinsic forces between particles are steric, and the
entropic effects of shape (which we term ``shape entropy''\footnote{The term
shape entropy has been used previously in unrelated contexts in
\cite{shapeent1,shapeent2,shapeent3}.}), can be isolated. Those works show
that shape entropy begins to be important when systems are at moderate density.
\cite{kamien}

In laboratory systems, however, it is not possible to isolate shape entropy
effects with as much control, and so the role of shape entropy in experiment is
less clear.  However, intuition suggests that shape entropy becomes important
when packing starts to dominate intrinsic interactions, and therefore should be
manifest in crowded systems in the laboratory.

Unlike other interactions, shape entropy is an emergent \footnote{We use
``emergent'' to denote observed macroscopic behaviors that are not seen in
isolated systems of a few constituents. At low pressures hard particles behave
like an ideal gas. Ordered phases only arise at reduced pressures of order one.}
quantity that is expected to become important as systems become crowded.
Although entropy driven phase behavior, from the crystallization of hard spheres
\cite{kirkwood,kirkwoodboggs1,kirkwoodboggs2,kirkmaunalder,alder,wood}, to the
nematic transition in hard rods \cite{onsager}, to colloid-polymer depletion
interactions \cite{ao}, has been studied for decades, linking microscopic
mechanisms with macroscopic emergent behavior is difficult in principle
\cite{moreisdifferent}. Hence, even for idealized systems, despite the
overwhelming evidence \textit{that} shape entropy is implicated in phase
behavior, understanding \textit{how} shape entropy is implicated is only now
starting to be distilled.
\cite{freons,strlekfresph,schillingpent,pped,fichthornrect,amirnature,amirtetphdiag,
dijkstraentdr,escobedo,zhaosquare,amirprl,rossi,trunctet,zoopaper,escobedordsq,geissleryang,
dijkstrasuperballs,nanooct,escopoly,dijkstracube,dijkstraplatelet,dijkstrabowl,
dijkstrabowl2,dijkstranonconvex,dftpoly,dijkstratcube,amirtbp,
kayliedef,youngmirkin,glotzermurray,epp,archimedean,bolestalapin}.
For example, the phase behavior of binary hard sphere mixtures
\cite{sandersbhs,murraysandersbhs,bartlettbhs,eldridgebhs,trizacbhs} or polygons
\cite{superdiskpack,zhaosquare,escobedordsq} can be deduced from global packing
arguments, but for many other shapes,\cite{zoopaper} including ``simple''
platonic solids like the tetrahedron \cite{amirnature} and its
modifications,\cite{trunctet} this is not the case.

One suggestion of how shape entropy is implicated in the phase behavior of
systems of anisotropic particles is through the idea of directional entropic
forces (DEFs).\cite{trunctet} Damasceno \textit{et al.} inferred the existence of
these forces by observing that in many idealized systems of convex polyhedral
shapes, one tends to observe a high degree of face-to-face alignment between
particles in crystals. However, the origin and strength of these forces are
unclear.

Here we use computer simulations to address how these forces arise, and
construct a rigorous theoretical framework that enables this investigation. Our
key results are: (i) We quantify pairwise DEFs in arbitrary systems, compute
them directly in several example systems, and show they are on the order of a
few $k_\mathrm{B}T$ just before the onset of crystallization. (ii) We show that
the microscopic mechanism underlying the emergence of DEFs is the need for
particles to optimize their local packing in order for the system to maximize
shape entropy. (iii) By computing quantities for DEFs that can be compared to
intrinsic forces between particles, we determine when shape entropy is important
in laboratory systems, and suggest how to measure DEFs in the lab. (iv) We
explain two notable features of the hard particle literature: the observed
frequent discordance between self-assembled and densest packing structures,
\cite{amirprl,trunctet,geissleryang,zoopaper,escobedo,amirtbp} and the high
degree of correlation between particle coordination in dense fluids and
crystals.\cite{zoopaper} (v) As we illustrate in Fig.\ \ref{fig-scope}, we show
that the same local dense packing mechanism that was known to drive the phase
behavior of colloid-depletant systems also drives the behavior of monodisperse
hard particle systems, thereby allowing us to view -- within a single framework
-- the entropic ordering considered here and in previous works
\cite{kirkwood,kirkwoodboggs1,kirkwoodboggs2,kirkmaunalder,alder,wood,onsager,
ao,depletion,kamien,freons,strlekfresph,schillingpent,pped,fichthornrect,amirnature,
amirtetphdiag,dijkstraentdr,escobedo,zhaosquare,amirprl,rossi,trunctet,zoopaper,
geissleryang,
dijkstrasuperballs,nanooct,escopoly,dijkstracube,dijkstraplatelet,dijkstrabowl,
dijkstrabowl2,dijkstranonconvex,dftpoly,dijkstratcube,amirtbp,
kayliedef,youngmirkin,glotzermurray,epp,archimedean,bolestalapin,
mason,zhaomason1,zhaomason2,stroock1,stroock2,yakerough,synderrough,
kraftetal,enttorque,lockkeyent,lockkey,pacman,pacsuprmol,fichthorn,
dogic}.

\section{Methods} \label{methods}
To compute statistical integrals we used Monte Carlo (MC) methods. For purely
hard particle systems, we employed single particle move Monte Carlo (MC)
simulations for both translations and rotations for systems of $1000$ particles at
fixed volume. Polyhedra overlaps were checked using the GJK
algorithm\cite{gjk} as implemented in~\cite{zoopaper}. For penetrable hard
sphere depletant systems we computed the free volume available to the depletants
using MC integration.

We quantified DEFs between anisotropic particles at arbitrary density using the
potential of mean force and torque (PMFT). Such a treatment of isotropic
entropic forces was first given by De Boer \cite{deboer} using the canonical
potential of mean force \cite{onsagerpmf}; for aspherical particles, first steps
were taken in this direction in~\cite{cole,croxtonosborn}.

Consider a set of arbitrary particles, not necessarily identical. Take the
positions of the particles to be given by $q_i$, and the orientations of the
particles to be given by $Q_i$. The partition function in the canonical
ensemble, up to an overall constant, is given by
\begin{equation}
  Z = \int [dq] [dQ] e^{-\beta U(\{q\},\{Q\})} \; .
\end{equation}
where $U$ is the potential energy for the interaction among the particles.
Suppose that we are only interested in a single pair of particles, which we
label with indices $1$ and $2$, and denote all of the other (sea) particles with a
tilde. The partition function is formally an integral over all of the
microstates of the system, weighted by their energies. Because we are interested
in a pair of particles, we do not want to perform the whole integral to compute
the partition function. Instead we break up the domain of integration into
slices in which the relative position and orientation of the pair of particles
is fixed; we denote this by $\Delta \xi_{12}$.

We choose to work with coordinates $\Delta \xi_{12}$ that are invariant under
translating and rotating the pair of particles. In two dimensions, a pair of
particles has three scalar degrees of freedom. In three dimensions, for generic
particles without any continuous symmetries, a pair of particles has six scalar
degrees of freedom. In the appendix \ref{Smethods-an} we give an
explicit form for these coordinates that is invariant under translations and
rotations of the particle pair, and the interchange of their labels. Note that
particles with continuous symmetry (\ie\ spherical or axial) have fewer
degrees of freedom. Separating out the integration over the sea particles gives
the partition function as
\begin{equation}\label{ZCAL}
  \begin{split}
    Z =& \int d\Delta\xi_{12} J(\Delta \xi_{12})
    e^{-\beta U(\Delta\xi_{12})}\\
    &\int [d\tilde{q}] [d\tilde{Q}]
    e^{-\beta U(\{\tilde{q}\},\{\tilde{Q}\},\Delta \xi_{12})} \; .
  \end{split}
\end{equation}
We formally integrate over the degrees of freedom of the sea particles to write
\begin{equation}\label{tildeFdef}
  Z = \int d\Delta\xi_{12} J(\Delta \xi_{12})
  e^{-\beta U(\Delta\xi_{12})} e^{-\beta \tilde{F}_{12}(\Delta \xi_{12})} \; ,
\end{equation}
where $\tilde{F}_{12}$ encodes the free energy of the sea particles with the
pair of interest fixed, and $J$ is the Jacobian for transforming from the
absolute positions and orientations of the particles in the pair to their
relative position and orientation. We define the PMFT for the particle pair
$F_{12}$ implicitly through the expression
\begin{equation}\label{Fdef}
  Z \equiv \int d\Delta\xi_{12} e^{-\beta F_{12}(\Delta\xi_{12})} \; .
\end{equation}
Equating the logarithms of the integrands on the right-hand sides of
Eqs.~\eqref{tildeFdef} and \eqref{Fdef} gives an expression for the PMFT
($F_{12}$)
\begin{equation}\label{PMFT}
  \begin{split}
    \beta F_{12}(\Delta\xi_{12}) = &
    \beta U(\Delta\xi_{12}) - \log J(\Delta \xi_{12})\\
    &+\beta \tilde{F}_{12}(\Delta \xi_{12}) \; .
  \end{split}
\end{equation}
In the \ref{Smethods-an} we outline how to extract the forces and torques from
Eq.~\eqref{PMFT}, and give example calculations that determine the thermally
averaged equations of motion for pairs of particles.
\begin{figure*}
  \begin{center}
  \includegraphics[width=12cm]{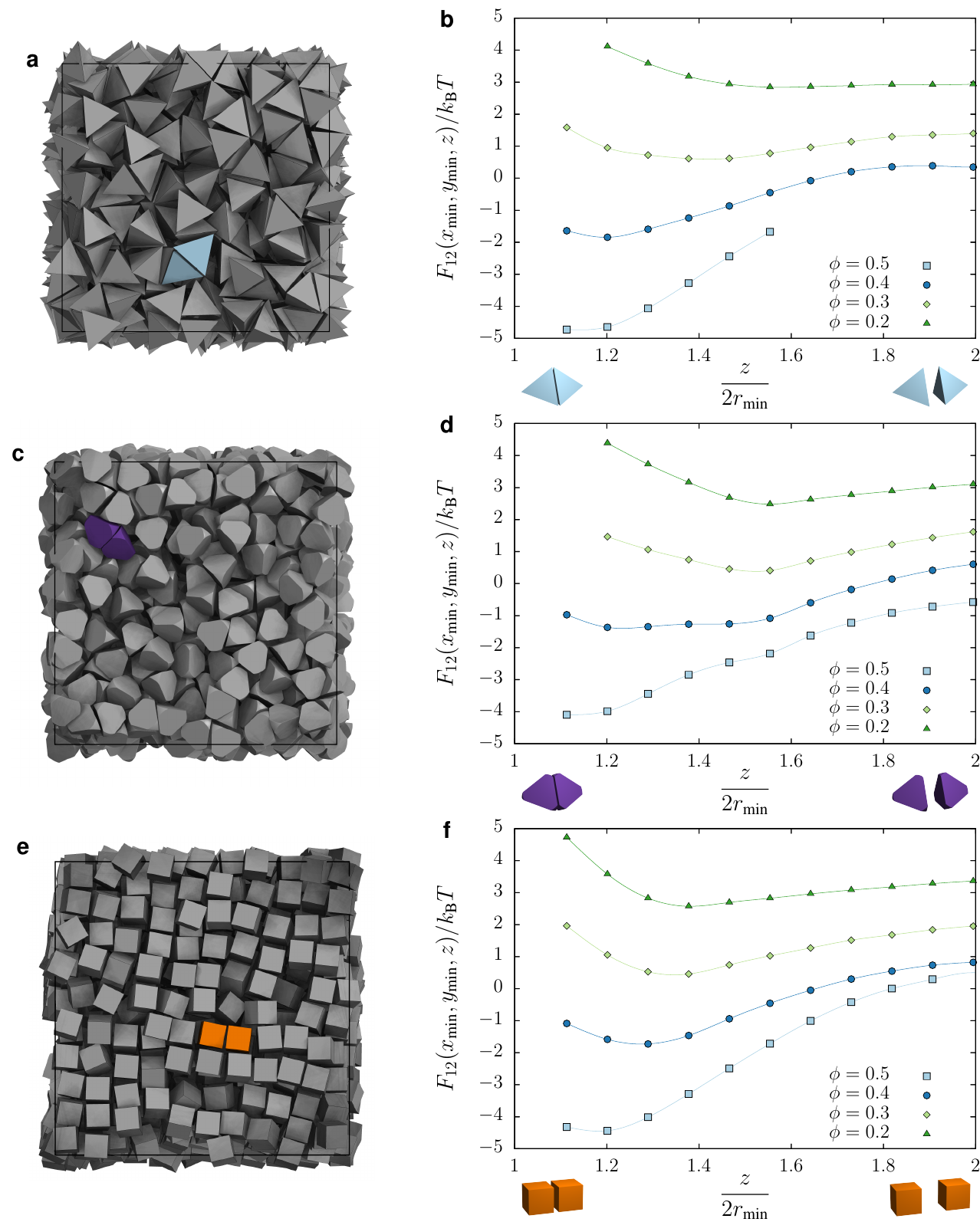}
  \end{center}
  \caption{
  In monodisperse systems, we compute the PMFT by considering pairs of particles
  (panels a, c, and e). Density dependence of the PMFT along an axis
  perpendicular to the polyhedral face for a hard tetrahedra fluid (b), a fluid
  of tetrahedrally faceted hard spheres (d), and a hard cube fluid (f). Data
  are computed from the frequency histogram of the relative Cartesian
  coordinates of pairs of particles in MC simulations of monodisperse hard
  particles, and correspond to the integration of the PMFT over relative
  orientation. We plot along the axis that contains the global minimum of the
  potential, and only plot data that are within $4\; k_\text{B}T$ of the global
  minimum at each respective density, as we are able to sample such points
  reliably. The perpendicular distance $z$ is given in units of the minimum
  separation between particles, which is twice the radius of the inscribing
  sphere of the given polyhedron. Error bars are smaller than the markers
  indicating data points, and a smooth curve through the data points in each
  series is used to guide the eye.
  \label{phidep_perp}}
\end{figure*}

In cases in which the particle pair of interest has only excluded volume
interactions, as in the remainder of this paper, it is convenient to combine the
first two terms to cast expression Eq.~\eqref{PMFT} as
\begin{equation}\label{PMFThard}
  \begin{split}
    F_{12}(\Delta \xi_{12}) =&
    -k_\text{B}T\log\left(H\left(d(\Delta \xi_{12})\right)
    J(\Delta \xi_{12})\right)\\
    &+\tilde F_{12}(\Delta \xi_{12})
  \end{split}
\end{equation}
where $H$ is the Heaviside step function that we use as a bookkeeping device to
ensure that the effective potential is infinite for configurations that are
sterically excluded, and $d(\Delta \xi_{12})$ is the minimum separation distance
of the particle pair in their relative position and orientation, which is
negative when the particles overlap, and positive when they do not.

When the sea particles are penetrable hard sphere depletants, \ie\ the sea
particles are an ideal gas with respect to each other but hard with respect to
the pair, the contribution of the sea particles to Eq.~\eqref{PMFT} can be
evaluated directly. If we have $N$ penetrable hard sphere depletants, we
evaluate the sea contribution $\tilde{F}_{12}(\Delta \xi_{12})$ in Eq.\
\eqref{PMFT} to be
\begin{equation} \label{FtwIdeal}
  e^{-\beta \tilde{F}_{12}(\Delta \xi_{12})}
  \propto V_\text{F}(\Delta \xi_{12})^N \; ,
\end{equation}
where $V_\text{F}$ is the free volume available to the sea particles. If we
consider two nearby configurations, we have that
\begin{equation} \label{dFIdeal}
  \begin{split}
  \beta(F_{12}'-F_{12})=& \beta(U'-U)-
  N\log\left(\frac{V_\text{F}'}{V_\text{F}}\right)\\
  &-\log J'+\log J\\
  \approx& \beta(U'-U)-\beta P(V_\text{F}'-V_\text{F})\\
  &-\log J'+\log J \; ,
  \end{split}
\end{equation}
where we have used the ideal gas equation of state for the depletant particles.
Thus, up to an irrelevant additive constant,
\begin{equation} \label{Fideal}
  \begin{split}
    \beta F_{12}(\Delta \xi_{12})=&\beta U(\Delta \xi_{12})
    -\beta P V_\text{F}(\Delta \xi_{12})\\
    &-\log J(\Delta \xi_{12}) \; .
  \end{split}
\end{equation}
The treatment of mixed colloid-polymer depletion systems as mutually hard
colloids in the presence of non-interacting polymers is known in the literature 
as the penetrable hard sphere limit \cite{depletion}. Hence Eq.\ \eqref{Fideal}
is the generalization of the Asakura-Oosawa~\cite{ao} result for depletion
interactions between spherical particles to particles of arbitrary shape.

\section{Results} \label{results}
\subsection{Entropic Forces in Monodisperse Hard Systems} \label{mono}
As argued in~\cite{trunctet,zoopaper} we expect that for a pair of polyhedra
the DEFs favor face-to-face arrangements. In terms of the PMFT, we therefore
expect face-to-face configurations to have the deepest well of effective
attraction.

We compute the force components of the PMFT in Cartesian coordinates for
polyhedra or polyhedrally faceted spheres and integrate over the angular
directions, as described in the appendix \ref{Smethods-num}. Also, appendix
\ref{Sresults} gives explicit results for torque components in an example
monodisperse hard system. We then use a set of orthogonal coordinate systems for
each face of the polyhedron (see Fig.\ \ref{tetcoord} for a schematic diagram of
this for a tetrahedron) and linearly interpolate the PMFT to the coordinate
frame of each facet.

In Fig.~\ref{phidep_perp} we plot the PMFT in the direction perpendicular to the
face for the three systems of: (a) hard tetrahedra (b) hard tetrahedrally
faceted spheres and (c) hard cubes at various densities shown in the legend.
We have chosen an axis that passes through the global minimum of the potential.
Several independent runs were averaged to obtain these results.
Because we are free to shift the PMFT by an additive constant, we have shifted the
curves for each density for clarity. We only show points that lie within $4\;
k_{B} T$ of the global minimum because we can sample accurately
at these points. As the density increases, the first minimum of the potential
decreases, and gets closer to contact. This indicates that the particles exhibit
greater alignment at higher densities as expected. Note that this is not merely
an artifact of the decrease in average particle separation at higher densities
because, as we show below and in appendix \ref{Sresults}, the alignment effect
is concentrated near the center of the facet.
\begin{figure*}
  \begin{center}
    \includegraphics[width=17cm]{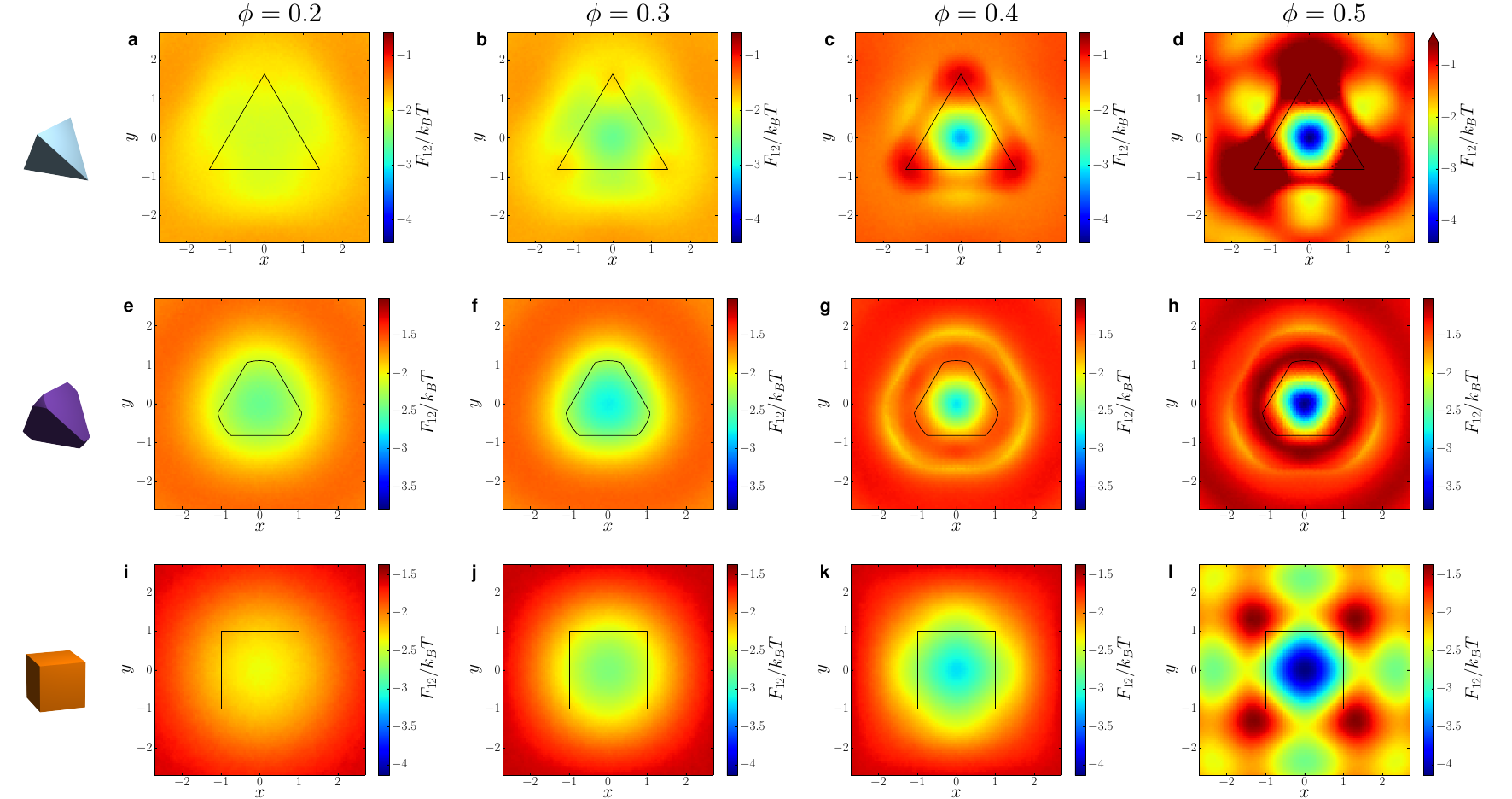}
  \end{center}
  \caption
  {
  Directional entropic forces are emergent in systems of particles
  and, as such, cannot be directly imaged through microscopy. Here we show the
  location of potential wells by taking slices of the PMFT (computed from the
  frequency histogram of the relative Cartesian coordinates of pairs of
  particles in MC simulations of monodisperse hard particles) parallel to the
  faces of a tetrahedron (a-d), a tetrahedrally faceted sphere (e-h), and a
  cube (i-l) at various packing fractions $\phi=0.2,\, 0.3,\, 0.4,\, 0.5$,
  indicated at the head of the column. As the packing fraction increases from
  left to right, the potential well becomes stronger, and its shape becomes more
  well-defined.
  \label{eppic}
  }
\end{figure*}

\begin{figure*}
  \begin{center}
    \includegraphics[width=17cm]{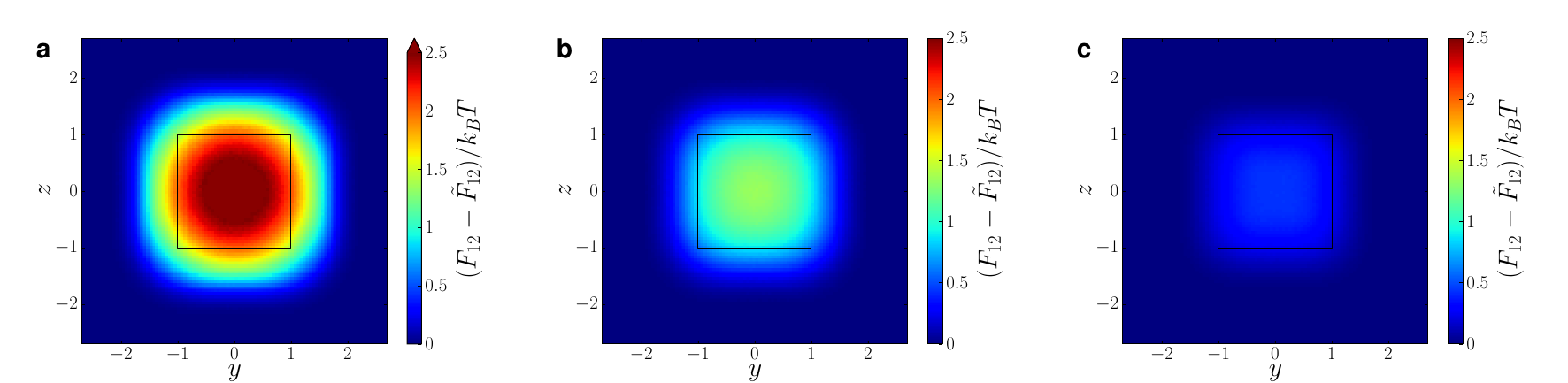}
  \end{center}
  \caption{
  The density independent contribution to the PMFT for cubes at distances of
  $z/2r_0 \approx 1.1$ (a), $1.2$ (b), and $1.3$ (c), which arises intrinsically
  from the pair of colloids under consideration, and contributes the first term
  factor on the right hand side in Eq.\ \eqref{PMFThard}. This contribution
  provides an effective repulsion between free particles, which is stronger at
  small separations (a) than larger ones (b) and (c). The pattern of the
  repulsion is opposite that of the overall PMFT between cubes seen in Fig.\
  \ref{eppic}, where both factors in Eq.\ \eqref{PMFThard} have been taken
  together.
  \label{pairfact}
  }
\end{figure*}

To understand what the PMFT is surrounding each shape we do the following. In
Fig.~\ref{eppic} we plot the PMFT in the plane parallel to the face that passes
through the global minimum of the potential (which can be identified by the
minimum of the respective curve in Fig.~\ref{phidep_perp}). The outline of one
of the faces of one particle of the pair is indicated by the solid line. The
second reference particle is allowed to have its center of mass anywhere on this
plane. The orientation of the second particle can vary and we integrate over all
the angles. Because the particles cannot overlap, in practice not all
orientations are allowed at close distances (we plot this effect for cubes in
Fig.~\ref{pairfact}). Each row shows (from left to right) increasingly dense
systems of tetrahedra (top row), tetrahedrally faceted spheres (middle row), and
cubes (bottom row), respectively.

Directional entropic forces originate from entropic patch sites \cite{epp} --
geometric features that facilitate local dense packing -- but, as emergent
notions, these patch sites cannot be imaged as, say, sticky patches created
through gold deposition on the surface of a nanoparticle can through electron
microscopy. Instead, in Fig.~\ref{eppic} we plot the location and strength of
entropic patches at different densities. These plots show that the entropic
patches lie at the centers of the facets. The effect of the pressure of
the sea particles can be seen by comparing the density dependent PMFT from Eq.\
\eqref{PMFThard}, plotted for cubes in Fig.~\ref{eppic}i-l, to the density
independent pair contribution for cubes in Fig.~\ref{pairfact}. The pair
contribution clearly drives pairs of cubes away from direct face-to-face
contact, and it is the other, contribution from the sea particles that
is responsible for driving them together. For the polyhedral shapes, the
coordination of the patches corresponds to the locations of the vertices of the
dual polyhedron, and the shapes of the patches themselves at high densities
appear to reflect the symmetry of the dual. In~\cite{epp}, we show that
systematic modification of the particle shape induces DEFs between particles
that lead to the self-assembly of target crystal structures.

Note that though the tetrahedron and the tetrahedrally faceted sphere share the
same point group symmetry, and the geometrical coordination of basins of
attraction is the same in both cases, the shape of the effective potential and
its strength are different. For example, contrasting the two shapes at
$\phi=0.4$ (see Fig.~\ref{eppic}) we see that the potential difference between
the center of the facet of the tetrahedrally faceted sphere and the truncated
vertex is more than a $k_\text{B}T$ different than in the case of the actual
tetrahedron. This indicates that small changes in particle shape can have
dramatic effects on the structural coordination of the dense fluid.

\begin{figure}
  \begin{center}
    \includegraphics[width=9.0cm]{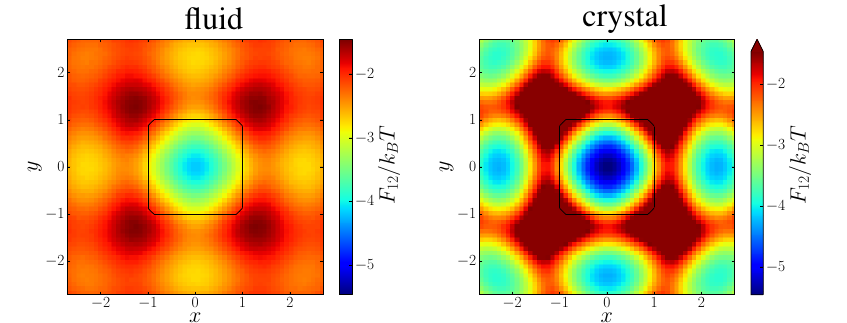}
  \end{center}
  \caption{
  Directional entropic forces emerge in the fluid phase (left) and persist into
  the crystal (right). Here we plot the PMFT for cubically faceted spheres
  (from the family of shapes studied in \cite{epp}), that are nearly perfect
  cubes. In the dense fluid at $\phi=0.5$ (left panel) and the crystal
  $\phi=0.6$ (right panel) the PMFT has a similar form, but the strength of the
  interaction is at least $1.5 \; k_\text{B}T$ stronger in the crystal.
  \label{pmftxtal}
  }
\end{figure}

The existence of the directionality in the PMFT in the dense fluid
is strongly suggestive that DEFs provide the mechanism for crystallization.  In
Fig.~\ref{pmftxtal} we show that DEFs persist, and increase, in the crystal. For
concreteness we study systems of cubically faceted spheres that are very close
to perfect cubes at a packing fraction of $\phi=0.5$ (fluid) and $\phi=0.6$
(crystal). As we show in \cite{epp}, at sufficiently high packing fractions,
these particles self-assemble a simple cubic lattice. Upon increasing the
packing density from the fluid to the crystal, the PMFT develops stronger
anisotropy. For example, by comparing the PMFT at the center of the facet to the
location of the vertex of the faceting cube, we note the difference between
identical points can increase by more than $1.5 \; k_\text{B}T$.

\subsection{Entropic Forces with Penetrable Hard Sphere Depletants} \label{phs}
We study DEFs as a function of colloid shape in systems with traditional weakly
interacting, small depletants. One system consists of a pair of spherical
particles that are continuously varying faceting with a single facet, in order
to promote locally dense packing. The other is a system of spherocylinders of
constant radius that are continuously elongated. In each case the alteration
creates a region on the surface of the particle with reduced spatial curvature.
\begin{figure*}
  \begin{center}
    \includegraphics[width=15cm]{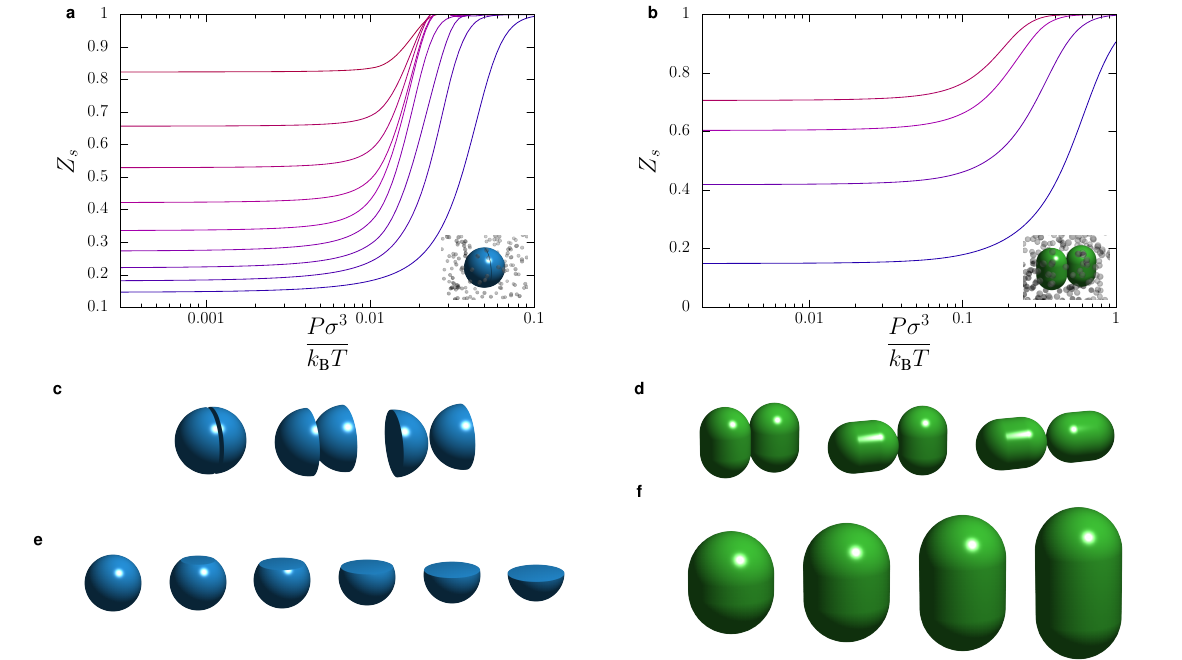}
  \end{center}
  \caption{The probability of specific binding, $Z_s$ for (a) particles with
  a single facet, and (b) capsules, in a bath of depletants as a function of
  depletant pressure. Panels (c) and (d) depict configurations that correspond
  to specific, semi-specific, and non-specific binding according to whether the
  binding occurs at sites of low curvature. Panels (e) and (f) show singly
  faceted spheres and spherocylinders with various facet and cylinder sizes,
  respectively. Curves in (a) and (b) show the probability of specific binding
  at various faceting amounts; with increasing patch size (more faceting) the
  color goes from blue to red.
  \label{depletionf}}
\end{figure*}

As the amount of alteration to particle shape increases, it leads to stronger
attraction between the sites of the reduced curvature, as encoded in the
probability of observing particles with these entropic patches \cite{epp}
adjacent.  In the case of the faceted particle, we vary the depth of the facet
linearly between zero (a sphere) and unity (a hemisphere), with the radius of
the sphere fixed to $10$ (see Fig.~\ref{depletionf}e). In the case of
spherocylinders, we studied cap radii fixed to $5$ and cap centers interpolating
linearly between $1$ (nearly spherical) and $4$ (an elongated spherocylinder),
where all lengths are in units of the depletant radius (see
Fig.~\ref{depletionf}f). See appendix \ref{Smethods-num} for computational
details. In Fig.~\ref{depletionf} a and b we show the probability that if a pair
of particles is bound, then they are bound patch-to-patch entropically (specific
binding) as a function of depletant pressure. Specific binding is depicted in
the inset images and the left hand particle pairs in panels c and d, and is
contrasted with patch-to-non-patch (semi-specific binding), and
non-patch-to-non-patch (non-specific binding) in the center and right hand
particle images in panels c and d. Panels a and b show that we can tune binding
specificity by adjusting the patch size and the depletant pressure.  Different
curves correspond to increasing patch size (more faceting) as the color goes
from blue to red.

\begin{figure}
  \centering
  \includegraphics[width=8.6cm]{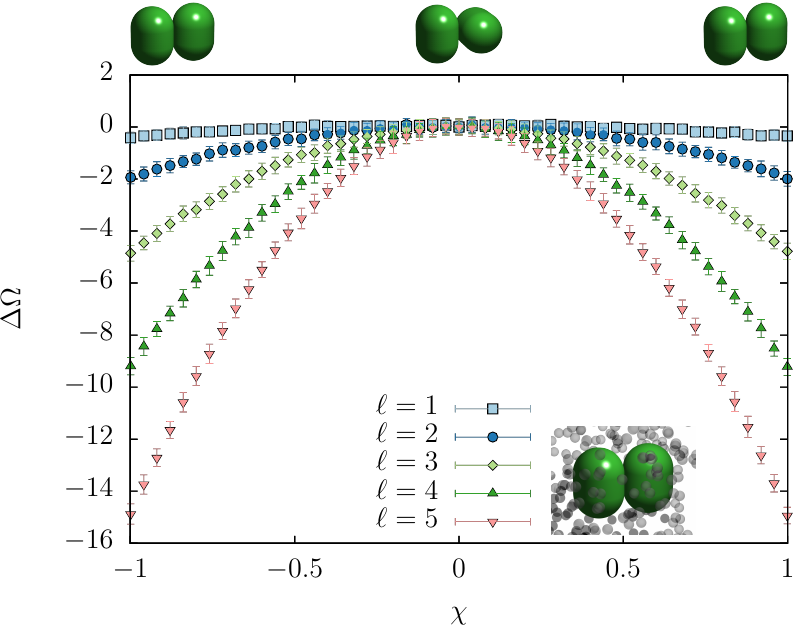}
  \caption{The angular ($\chi$) dependence of the portion of the PMFT directly
  attributable to modification of particle shape ($\Omega$, see
  Eq.\ \eqref{omegadef}, normalized to $0$ at $\chi=0$) for
  spherocylinders in the presence of penetrable hard sphere depletants for
  different cylinder side lengths. For short side length ($\ell=1$, light blue
  squares) there is very little angular variation indicating very little
  entropic torque. As the side length, $\ell$, increases from to $5$ (in units
  of the depletant radius, pink triangles), the strength of the dependence on
  $\chi$ increases, indicating greater torques, as expected.
  \label{torquef}}
\end{figure}

For spherocylinders, it is straightforward to show the effect of the entropic
patches in generating torques that cause the particles to align. To isolate the
part of the torque that comes from the patch itself, we rearrange
Eq.~\eqref{Fideal} for non-overlapping spherocylinders to get
\begin{equation} \label{Ftorque}
  \frac{F_{12}+k_\text{B} T \log J}{P\sigma^3} = -\frac{V_F}{\sigma^3}
\end{equation}
Note that, conveniently, for ideal depletants, the expression on the right-hand
side is independent of the depletant pressure. For clarity, we fix the
separation distance $R$ between the spherocylinders centers of mass to be $1\%$
larger than the spherocylinder diameter (which is the minimum separation
distance), and fix the orientation of each spherocylinder to be normal to the
separation vector between them ($\phi_1=\phi_2=0$ in the coordinates in appendix
\ref{Smethods-an}).
Because it is always possible to shift the PMFT by a constant, for normalization
purposes we define
\begin{equation} \label{omegadef}
  \Omega(\chi) \equiv -\frac{V_F(R,\chi)}{\sigma^3}
\end{equation}
where $\chi$ describes the angle between the spherocylinder symmetry axes.
In Fig.~\ref{torquef} we plot $\Delta\Omega \equiv \Omega(\chi)-\Omega(0)$ for
spherocylinders of different aspect ratios. For small side lengths of the
spherocylinder, there is very weak dependence of the PMFT on $\chi$. As the
length of the cylinder increases (and therefore as the entropic patch gets
larger) the $\chi$ dependence of the PMFT becomes more pronounced. This means
that not only do the particles coordinate at their entropic patch sites,
but there also exists a torque \cite{enttorque} that aligns the patches.

\section{Discussion}\label{discuss}
\subsection{PMFT as an Effective Potential at Finite Density}
Extracting physical mechanisms from systems that are mainly governed by entropy
is, in principle, a difficult task. This difficulty arises because entropic
systems have many degrees of freedom that exist at the same energy or length
scale. Typically, disparities in scale are used by physicists to construct
``effective'' descriptions of physical systems, in which many degrees of freedom
are integrated out and only the degrees of freedom key to physical mechanisms
are retained. Determining which degrees of freedom are key in entropic systems
is difficult because of the lack of natural hierarchies, but to extract physical
mechanisms it is still necessary.

To extract the essential physics of shape for anisotropic colloidal particles,
we computed the PMFT. In so doing, we described the PMFT as an effective
potential, but we emphasize here that it is an effective potential in a
restricted sense of the term. The restriction comes about because when we
separate a system into a pair of reference colloids plus some sea particles, the
sea particles can be identical to the reference colloids. Only the pair of
reference colloids is described by the effective potential, which means \emph{it
describes the behavior of a system of only two particles with the rest
implicit}. Indeed, the rest of the colloids in the system are treated as an
``implicit solvent'' for the pair under consideration. Because of this fact, the
PMFT for monodisperse colloids is not synonymous with the bare interaction
potential for the whole system.  Note that in systems of multiple species, there
is some possibility of a broader interpretation of the PMFT for a single species
having integrated out another species \cite{dijkstrabhs}, but even in that case,
difficulties arise.\cite{beware} \Eg, Onsager treated the nematic transition in
hard rods by considering rods of different orientations as different ``species''
\cite{onsager}, but an effective potential between rods of the same ``species''
or orientation does not capture the physics of the system in the same way that
an effective potential between rods induced by polymer depletants
would.\cite{kamien}

Interpreting the PMFT from Eq.\ \eqref{PMFT} in the restricted sense, we see
that it naturally exhibits three contributions. (i) The first term on the right
hand side is the bare interaction between the pair of particles, which
originates from van der Waals, electrostatic, or other interactions of the
system of interest. It encodes the preference for a pair of particles to be in a
particular relative position and orientation. Here, we are mostly concerned with
hard particles, so this term encodes the excluded volume interaction. (ii) The
second term on the right hand side is the logarithmic contribution from the
Jacobian; it counts the relative number of ways that the pair of particles can
exist in a relative position and orientation. For a more technical discussion of
this term, see appendix \ref{Sdiscuss}. (iii) The third term is the free energy
of the sea particles that are integrated out, keeping the relative position and
orientation of the pair fixed. In purely hard systems this last term is sea
particle entropy, which is just the logarithm of the number of microstates
available to the sea particles for that configuration of the pair.

The free energy minimizing configuration of the pair of particles is determined
by a competition between the three terms in Eq.\ \eqref{PMFT}. Consider the
contribution from the third term. For any non-attractive bare interactions
between sea particles, if the pair of particles is separated by a distance that
is less than the effective diameter of the sea particles\cite{striperepuls}, and
adopts a configuration that packs more densely, the free energy of the sea
particles will never increase.  This is the case for bare interactions that are
repulsive due to excluded volume (in which case the system is dominated by
entropy) and for sea particles that are soft and interpenetrable.

In the absence of the first two terms in Eq.\ \eqref{PMFT}, the third term will
drive the pair of particles into denser local packing configurations. The
driving force for local dense packing becomes stronger as the system density
increases. To understand why this occurs, we note that DEFs arise by the pair of
particles balancing the pressure of the sea particles. Rather than considering the
effects of the sea on the pair, we will momentarily consider the effects of the
pair on the sea. The pair of interest forms part of the `box' for the sea
particles. Density dependence arises because changes in the relative position
and orientation are related to the local stress tensor. This contribution is on
the order of $P\sigma^3/k_\mathrm{B}T$, where $P$ is the characteristic scale of
the local stress tensor, which we expect to be related to the pressure, and $\sigma$
is a characteristic length scale. Since this contribution is density dependent
the effective forces between particles are emergent. These emergent forces are
directional, because some local packing configurations are preferred over
others. For example, dense local packing configurations involve face-to-face
contact for many particles, which is why face-to-face contact is observed in so
many hard-particle crystals, and is borne out in Fig.~\ref{eppic}. This is also,
of course, akin hard rods aligning axes in nematic liquid
crystals\cite{onsager}. Indeed, in repulsive systems, only at sufficiently high
densities is the system able to overcome any repulsion from the first two terms
in Eq.\ \eqref{PMFT}, plotted in Fig.\ \ref{pairfact} for cubes, which drive the
system away from locally dense packings. \emph{It is this competition between
the drive towards local dense packing supplied by the sea particles' entropy,
and the preferred local relative positions and orientations of the pair, which
is induced by the maximization of shape entropy, that determines the
self-assembly of entropic systems of repulsive (hard) shapes.}

Two general lessons from the literature of hard particle self-assembly are
intuitive in light of the mechanism we have described.

First, in a previous systematic study of shape and self-assembly using a
family of highly symmetric convex polyhedra,\cite{zoopaper} it was shown that
there is a remarkable correlation between the number of nearest neighbors in
the crystal structure of a given polyhedron, and the number of nearest
neighbors in the dense fluid. Since systems at finite density exhibit a drive
towards local dense packing, we would expect that the organization of neighbor
shells to be determined by the same local packing considerations in both the
dense fluid and the crystal. Indeed, we might expect to predict crystal
structure based on local packing considerations alone. In recent work we
designed particle shape to favor certain local dense packing arrangements to
self-assemble targeted crystal structures.\cite{epp} We viewed the induced
anisotropy of the effective interactions between particles as the entropic
analogue of the intrinsic anisotropic interactions between enthalpically patchy
particles\cite{patchy,glotzsolomon}. Attractive entropic patches are features in
particle shape that facilitate local dense packing.

Second, hard particle systems often assemble into their densest
packing structure, however some do not
\cite{amirprl,trunctet,geissleryang,zoopaper,escobedo,amirtbp}. The present
work helps to highlight two key differences between self-assembly and packing.  (i)
Systems self-assemble because the second law of thermodynamics drives them free
energy minima, so like packing, self-assembly has a mathematical optimization
problem at its root. In assembling systems, all three terms in Eq.\ \eqref{PMFT}
contribute, but in the infinite pressure limit relevant for packing, the entropy
of the pair of particles does not contribute. This is because the first term
provides the steric hindrance between particles, and the third term should scale
with pressure, whereas the second term encodes the entropy of a set of relative
positions and orientations of the pair, and this factor becomes irrelevant in
the infinite pressure limit.  (ii) Assembly is related to \emph{local} dense
packing, that we have characterized here with two-body potentials, whereas
\emph{global} dense packing is ostensibly an ``all-body problem.'' Given these
two differences it might be more surprising that densest packing solutions ever
coincide with assembled structures, than the fact that they frequently do not.
However, the pair entropy term often only separates particles, while preserving
their alignment, \eg\ we showed particles are more frequently observed with
face-to-face alignment at small separations, than in direct face-to-face
contact. Moreover, in practice (\eg\ \cite{dfamilyp}), the solution to the
all-body packing problem is often given by the solution to a few-body problem.

\subsection{Unification} \label{scope}
We have seen above that directional entropic forces, as captured by the PMFT,
arise when particles maximize shape entropy through local dense packing. We
verified this drive to local dense packing with direct calculations for
monodisperse hard shapes, and for hard shapes in a sea of penetrable
hard sphere depletants. Monodisperse hard-particle systems and colloid-depletant
systems would, at first glance, seem to be different.

However, depletion systems\cite{ao} and monodisperse systems of hard anisotropic
shapes are both entropy-driven, as is the crystallization of hard
spheres,\cite{kirkwood,kirkwoodboggs1,kirkwoodboggs2,kirkmaunalder,alder,wood}
and the nematic liquid crystal transition in hard rods \cite{onsager}.  Indeed,
since original work by Asakura and Oosawa \cite{ao} on depletion interactions,
it has been well known that the depletion-induced aggregation of colloids arises
from an osmotic effect in which depletants liberate free volume for themselves
by driving the colloids into dense packing configurations. However, earlier work
of de Boer \cite{deboer} suggested that considerations of the potential of mean
force in systems of isotropic particles leads to pairs of particles having
certain preferred distances because they balance their own repulsive forces with
the sea particles' preference for them to pack more densely.

In traditional depletion systems, the depletants are small compared with the
colloids, spherical, and interpenetrable. In such systems, there is a
substantial literature on the anisotropic binding of colloids by depletants
starting with~\cite{mason}, and followed by work on rough
colloids~\cite{zhaomason1,zhaomason2,stroock1,stroock2,yakerough,synderrough,
kraftetal}, as well as more anisotropic
shapes~\cite{enttorque,fichthorn,rossi,youngmirkin,dogic}, and lock-and-key systems
\cite{lockkeyent,lockkey,pacman,pacsuprmol}. In all of these systems, the
anisotropic binding can be seen as resulting from a sea of depletants forcing
colloids to adopt local dense packing configurations.

However, one might ask how large, or hard, or aspherical can the depletants be,
before they cease to act as depletants? In the original work of
Asakura and Oosawa \cite{ao}, depletants were not restricted to be small; large
depletants are the so-called ``protein limit''\cite{depletion}. In experimental
systems, and accurate models, depletants are not freely interpenetrable.
\cite{baumgartl,dijkstrabhs,depletion} In the original work of
Asakura and Oosawa \cite{ao}, depletants were imagined to be polymers, which are
certainly not spherical, but were modelled as spheres.

The fact that depletants can be aspherical, or hard, or large, is of
course well-known; in all cases depletants exert an osmotic pressure that causes
colloids to aggregate, as the colloids adopt dense packing configurations.
However, we argued above that consideration of the three contributions to the
PMFT in Eq.\ \eqref{PMFT} leads to a drive towards local dense packing in
generic systems in which the sea particles can be \emph{simultaneously}
arbitrarily aspherical, hard, and large. Indeed the sea particles can be the
same species as the colloids, and still behave as ``depletants''. Hence the
classic works on the entropic behavior of systems as diverse as hard shapes
from
spheres\cite{kirkwood,kirkwoodboggs1,kirkwoodboggs2,kirkmaunalder,alder,wood},
to rods\cite{onsager}, to tetrahedra\cite{amirnature}, and colloid-depletant
systems\cite{ao,depletion} are not similar only because their phase behavior is
driven by entropy. \emph{In addition, they are similar because entropy controls
their phase behavior through a preference for local dense packing.}

In Fig.\ \ref{fig-scope} we showed a schematic representation of the family of
systems whose phase behavior is governed by the shape-entropy driven mechanism.
We represent the family of systems on a set of three axes.
One axis represents the shape of the colloidal pair of
interest, in which we think of spheres as occupying the ``zero'' limit and rods
infinity. A second axis is the softness of the sea particles; one limit
represents hard-core particles (vanishing softness) and the other an
interpenetrable ideal gas (for spherical sea particles, in the case that the sea
particles can be considered as depletants, this is the so-called penetrable hard
sphere limit \cite{depletion}). A third axis is the inverse size of the sea
particles, where a monodisperse system is at the origin, and the other limit has
sea particles being very small compared to the ``colloids''. One could also
include other axes, such as the shape of the sea particles, or the density of
the system, but we omit them for simplicity.

As a guide we have indicated the location of some well-studied model and
experimental systems on these axes. Hard spheres sit at the origin
(c),\cite{kirkwood,kirkwoodboggs1,kirkwoodboggs2,kirkmaunalder,alder,wood}
hard rods\cite{onsager} lie at the extreme of the shape axis (a), and
Asakura and Oosawa's model of depletants\cite{ao} lies in the plane of zero shape
(between d and e). Some depletion systems are indicated (i,k,m); however in
experiments, particles are not all of the same shape, and are sometimes actually
binary (j,l) or ternary (n) mixtures. Monodisperse hard particle systems lie
along the shape axis (b,f), but are simply a limit of binary hard systems (g,h).

\subsection{DEFs in Experimental Systems}
We have shown that DEFs arise from a mechanism that occurs in a wide variety of
idealized systems. Do DEFs and shape entropy matter for experimental systems?
Using systems in which shape entropy is the only factor that determines
behavior, we have shown that in several example systems DEFs are on the order
of a few $k_\mathrm{B}T$ around the onset of crystallization. This puts DEFs
directly in the relevant range for experimental systems. If in experiment there
is sufficient control over the intrinsic forces between colloids that puts them
at much less than the $k_\mathrm{B}T$ scale, systems will be shape entropy
dominated. If the intrinsic forces are on the order of a $k_\mathrm{B}T$, then
there will be a competition between directional entropic forces and intrinsic
forces. If the scale of the intrinsic forces are much more than a few
$k_\mathrm{B}T$, intrinsic forces will dominate shape entropy effects.

Can these forces be measured in the laboratory? In colloidal experiments,
effective interaction potentials can be inferred from the trajectories of the
colloids \textit{in situ}, \eg\ by confocal microscopy~\cite{intconf}.
Recently, it has become possible to image anisotropic colloids in confocal
microscopes and extract particle positions and
orientations.\cite{rodviz,benellimg} The determination of both particle
positions and orientations makes it possible to directly extract the PMFTs
computed here from experimental data. For isotropic potentials of mean force,
this investigation has already been carried out by analyzing the pair
correlation function $g(r)$.\cite{intconf} As in \cite{intconf}, we can
interpret this potential measurement as describing the thermally averaged time
evolution of the relative position and orientation of pairs of particles.
Furthermore, we have shown that pairs of particles are described by the PMFT
regardless of the properties of the sea particles, so the PMFT describes the
average time evolution of pairs of particles in monodisperse systems, as well as
colloid-depletant systems.

\section{Conclusions}\label{conclusions}
In this paper we showed that shape entropy drives the phase behavior of systems
of anisotropic shapes through directional entropic forces. We defined DEFs in
arbitrary systems and showed that they are emergent, and on the order of a 
few $k_\mathrm{B}T$ just below the onset of crystallization in example hard
particle systems. By quantifying DEFs we put the effective forces arising from
shape entropy on the same footing as intrinsic forces between particles that are
important for self-assembly. In nano- and colloidal systems our results show
that shape entropy plays a role in phase behavior when intrinsic forces between
particles are on the order of a few $k_\mathrm{B}T$ or less. This figure guides
the degree of control of intrinsic forces in particle synthesis required for
controlling shape entropy effects in experiment. In microbiological systems it
facilitates the comparison of the effects of rigid shape in crowded environments
with the elasticity of the constituents. We suggested how existing experimental
techniques could be used to measure these forces directly in the laboratory. We
showed that the mechanism that generates DEFs is that maximizing shape entropy
drives particles to adopt local dense packing configurations.

Finally, we demonstrated that shape entropy drives the emergence of DEFs, as
particles adopt local dense packing configurations, in a wide class of soft
matter systems. This shows that in a wide class of systems with entropy driven
phase behavior, entropy drives the phase behavior through the same mechanism
\cite{kirkwood,kirkwoodboggs1,kirkwoodboggs2,kirkmaunalder,alder,wood,onsager,
ao}.

\begin{acknowledgments}
We thank Randall Kamien and Daan Frenkel for helpful discussions
concerning the early literature, Alexander Grosberg for helpful discussions
concerning the interpretation of the PMFT, and for comments on an earlier
version of this work, and Ben Schultz for helpful suggestions. We are
grateful to Henk Lekkerkerker, Robert Evans, and Paul Chaikin for helpful
feedback and encouragement. This material is based upon work supported by, or
in part by, the U.S.\ Army Research Office under Grant Award No.\
W911NF-10-1-0518, the DOD/ASD(R\&E) under Award No.\ N00244-09-1-0062, and the
Department of Energy under Grant No.\ DE-FG02-02ER46000. The calculations for
hard particle systems was supported by ARO. D.K.\ acknowledges funding by the
FP7 Marie Curie Actions of the European Commission, Grant Agreement
PIOF-GA-2011-302490 Actsa. The PMFT derivation and depletion calculations were
supported by the Biomolecular Materials Program of the Materials Engineering
and Science Division of Basic Energy Sciences at the U.S. Department of Energy
under grant DE-FG02-02ER46000. Any opinions, findings, and conclusions or
recommendations expressed in this publication are those of the authors and do
not necessarily reflect the views of the DOD/ASD(R\&E).
\end{acknowledgments}

\appendix
\section{Supplementary Methods}\label{Smethods}
\subsection{Further Analytical Considerations}\label{Smethods-an}
\subsubsection{Jacobian Factor}\quad
We have included the Jacobian factor in our definition of the PMFT. The
motivation for doing this is that by explicitly including this term we do not
want to either ignore or introduce any artifacts that might stem from a poor
choice of coordinates for a given problem. Note that for general particles in
three dimensions, the PMFT is defined on the space
$\mathbb{R}^3\times\mathbb{RP}^3$. This means that even if the pair interaction
is purely hard, whenever there is non-trivial shape, some relative pair
configurations are preferred over others. In standard treatments of the
isotropic potential of mean force, it is not conventional to include the
Jacobian factor. The choice not to include it in that case is well-motivated by
the existence of a single ``natural'' coordinate system in that case that
prevents any ambiguity there. In either case, one must take account of the
inclusion or not of this factor in writing down the effective equations of
motion for the pair.

\subsubsection{Axisymmetric Coordinate System for PMFT}\quad
We present an explicit computation of the Jacobian of the change of variables
between the natural coordinates of a pair of particles, and the scalar invariant
quantities that describe any such pair. We do so in the simpler case of
axisymmetric particles. The general case can be computed straightforwardly in
the same manner, but the expressions are cumbersome.

We take the first particle to be at the origin, with its symmetry axis oriented
in the positive $z$ direction. Using the azimuthal symmetry with this placement,
we fix the second particle's position in the $xy$-plane, without loss of
generality to be along the $x$ axis. This gives the orientation of the second
particle as
\begin{equation} \hat n_2 = \sin\theta\cos\varphi \hat x +
  \sin\theta\sin\varphi \hat y + \cos\theta \hat z
\end{equation}
where $\theta$ and $\varphi$ are spherical coordinates in the coordinate system
of the second particle, and its position as
\begin{equation}
  r_2-r_1 = \rho \hat x + z \hat z
\end{equation}
where $\rho$ and $z$ are cylindrical coordinates in the first particle's
coordinate system. The volume form that appears in the integral that computes
the partition function is
\begin{equation}
  dV = \rho\sin\theta d\rho dz d\theta d\varphi
\end{equation}
Now we make the change of variables to the scalar invariant quantities by taking
\begin{equation}
  \begin{split}
    R&=\sqrt{z^2+\rho^2}\\
    \phi_1&=\frac{z}{\sqrt{z^2+\rho^2}}\\
    \phi_2&=-\frac{\rho\sin\theta\cos\varphi+z\cos\theta}{\sqrt{z^2+\rho^2}}\\
    \chi&=-\cos\theta
  \end{split}
\end{equation}
Inverting the relationship between the two coordinate
systems gives
\begin{equation}
  \begin{split}
    \rho &= R\sqrt{1-\phi_1^2}\\
    z &= R\phi_1\\
    \theta &= -\cos^{-1}\chi\\
    \varphi &= \cos^{-1}
    \left(\frac{-\phi_1\chi-\phi_2}{\sqrt{(1-\chi^2)(1-\phi_1^2)}}\right)
  \end{split}
\end{equation}
The computation of the determinant is simplified by noting the dependence of $\rho$
and $z$ on only $R$ and $\phi_1$, and $\theta$ on only $\chi$. This means that
we only need to consider the dependence of $\varphi$ on $\phi_2$. Taking the
determinant of this leads to the new volume form
\begin{equation}
  dV = \frac{R^2 dR d\phi_1 d\phi_2 d\chi}
  {\sqrt{1-\chi^2-\phi_1^2-\phi_2^2-2\phi_1\phi_2\chi}}
\end{equation}
In the absence of the excluded volume (or other) interaction between the
particles, this expression measures the volume of configuration space available
to a pair of free particles in a particular translationally and rotationally
invariant configuration.

To verify that our expression correctly encodes the density of states for two
free axisymmetric particles, we consider the following scenario. Suppose we had
a pair of axisymmetric particles each undergoing Brownian motion with the
constraint that their centers of mass could never be separated by a distance
greater than $R_\text{max}$, but that they were otherwise free to move,
including to interpenetrate. If we were to make some $N_\text{obs}$ uncorrelated
observations of the particles for each of which we determine the values of $R$,
$\phi_1$, $\phi_2$, and $\chi$, we would find that their frequency distribution
would converge to something proportional to the Jacobian of our coordinate
transformation in the limit that $N_\text{obs}\to\infty$. We therefore verified
our expression by performing precisely this calculation.

\subsubsection{General Coordinate System for PMFT}\quad
For the general case in three dimensions, a pair of particle has six degrees of
freedom that are invariant under global translations and rotations. Let us take
the particles to be situated at $\vec x_{1,2}$, and have orientations $\bq_{1,2}$.
Starting with the positions, the separation of the particles in position space
is simply given by the vector $\vec r_{12} = \vec x_2-\vec x_1$. This gives a
set of coordinates that are invariant under translations, but not under
rotations. We therefore seek to form six scalars by taking combinations of this
vector with the particle orientations.

To treat the particle orientations on a similar footing to the positions, we
need to determine the separation in orientation of the particles.
We note that that each of the particle orientations is given by an element of
$SO(3)$, the rotation group in three dimensions. While it is possible to work
with this group directly, the calculations that follow can be simplified greatly
by noting that $SU(2)$ is the double cover of $SO(3)$, and working with $SU(2)$
instead. For concreteness, we will use the conventions common in quantum
mechanics and write our particle orientations as rotation operators according to
\begin{equation} \label{qconv}
  \bq_i = e^{i\frac{\theta_i}{2} \hat n\cdot \vec\sigma} \; ,
\end{equation}
where $\theta_i$ is the angle of rotation $\hat n_i$ is the normal to the plane
of the rotation and $\vec \sigma$ are the Pauli matrices. From this form it is
clear that under a global rotation, the particle orientations transform in the
reducible representation expressed in Young tableaux as
\begin{equation} \label{2x2eq1p3}
  \yng(1)\otimes\yng(1) = \yng(1,1)\oplus\yng(2) \; .
\end{equation}
This can be understood intuitively in the following way: if you rotate a
particle by some amount in some plane, two different observers will agree on
the amount of the rotation, but will give the normal to the plane in their own
coordinates. The amount of the rotation is the scalar $\yng(1,1)$, and the
normal to the plane is the vector $\yng(2)$. In a similar fashion to the way in
which we combined particle positions to yield a relative position, we will
combine particle orientations. A Clebsh-Gordan decomposition of the particle
orientations gives
\begin{equation} \label{2x2x2x2}
  \begin{split}
    \left(\yng(1)\otimes\yng(1)\right)
    \otimes\left(\yng(1)\otimes\yng(1)\right) = \; & 
    \yng(2,2)\oplus\yng(2,2)\\
    &\oplus\yng(3,1)\oplus\yng(3,1)\oplus\yng(3,1)\\
    &\oplus\yng(4) \; ,
  \end{split}
\end{equation}
which means that the combination of the two orientations yields two scalars
(spin 0), three vectors (spin 1), and a tensor (spin 2).

For convenience, we note that the spin 1 representation of $SU(2)$ $\yng(2)$ is
also the adjoint representation of $SU(2)$. This means that we can write vectors
in ordinary space by using the Pauli matrices as Cartesian unit vectors
according to
\begin{equation} \label{vecadj}
  \vec v = \sum_i \vec v\cdot \hat e_i \sigma_i \; .
\end{equation}
In this representation, if we combine orientations expressed in terms of Pauli
matrices as in \eqref{qconv}, products of orientations that are proportional to
the identity matrix are scalar quantities, and products that are not are
vectors. We will use this fact momentarily.

For the purposes of
creating scalar invariants by combining them among themselves, and with the
particle separation in position, the scalars and vectors that arise from
combining orientations are of interest. We
determine these scalars and vectors explicitly by taking symmetrized Hermitian
products of the particle orientations.
We find the scalars can be expressed as
\begin{equation}
  \begin{split}
    S_{12} =& \frac12(\bq_1\bq_2^\dagger + \bq_2\bq_1^\dagger) \\
    U_{12} =& \frac14(\bq_1\bq_2+\bq_2\bq_1+\bq_1^\dagger \bq_2^\dagger
    + \bq_2^\dagger \bq_1^\dagger)
  \end{split}
\end{equation}
and the vectors as
\begin{equation}
  \begin{split}
    V_{12} =& \frac{i}{4}(\bq_1\bq_2-\bq_2\bq_1+\bq_1^\dagger \bq_2^\dagger-
    \bq_2^\dagger \bq_1^\dagger) \\
    W_{12} =& -\frac{i}{4}(\bq_1\bq_2+\bq_2\bq_1-\bq_1^\dagger
    \bq_2^\dagger- \bq_2^\dagger \bq_1^\dagger) \\
    T_{12} =& -\frac{i}{2}(\bq_1 \bq_2^\dagger - \bq_2 \bq_1^\dagger)
    - V_{12} \\
  \end{split}
\end{equation}
which we have identified according to their matrix form.
Using our convention for representing the particle orientations we find
\begin{equation}
  \begin{split}
    S_{12} =& \cos\left(\frac{\theta_1-\theta_2}{2}\right) \\
    U_{12} =& \cos\left(\frac{\theta_1+\theta_2}{2}\right) \\
    V_{12} =&
    \frac12(\hat{n}_1\times\hat{n}_2)(S_{12}-U_{12}) \\
    W_{12} =&
    \sin\left(\frac{\theta_1+\theta_2}{2}\right)
    \frac{\hat{n}_1+\hat{n}_2}{2}
    +
    \sin\left(\frac{\theta_1-\theta_2}{2}\right)
    \frac{\hat{n}_1-\hat{n}_2}{2}
    \\
    T_{12}
    =&
    \sin\left(\frac{\theta_1+\theta_2}{2}\right)
    \frac{\hat{n}_1+\hat{n}_2}{2}
    -
    \sin\left(\frac{\theta_1-\theta_2}{2}\right)
    \frac{\hat{n}_1-\hat{n}_2}{2}
  \end{split}
\end{equation}
By combining these quantities with $\vec r_{12}$ we can form the six scalar
invariants
\begin{equation}
  \{ |\vec{r}_{12}|, \; S_{12}, \; U_{12}, \;
  \hat{r}_{12}\cdot W_{12}, \; \hat{r}_{12}\cdot V_{12}, \;
  W_{12}\cdot T_{12} \}
\end{equation}

For the purposes of showing the coordination of particles in space at the
location of faces or facets, it is convenient to integrate over some of the
angular degrees of freedom, and to work in Cartesian coordinates. In particular,
it is very convenient to work in Cartesian coordinates adapted to the particle
facets as shown in Fig.~\ref{tetcoord} for tetrahedral facets.

\subsubsection{Forces and Torques}\label{fandt}\quad
For concreteness, we give an example for how to compute the forces and torques
from the PMFT. As usual, forces and torques arise from taking the negative
gradient of the potential. For force components this is straightfoward. For
torques, we give an explicit expression for the case of axisymmetric particles.

To compute the torques, we continue to work in terms of rotation matrices in the
spin $\frac12$ representation of $SU(2)$. If $\bq$ is the rotation, then to
determine the torque, we must differentiate the PMFT with respect to it. If we
represent the rotations in the canonical fashion, and use Pauli matrices as the
basis vectors of the Cartesian space coordinates, then we have scalar products
of the form
\begin{equation}
  \vec{a} \cdot \vec{b} = \frac12\Tr(a^\dagger b)
\end{equation}
and cross products of the form
\begin{equation}
  \vec{a} \times \vec{b} = \frac{1}{2i} [a,b]
\end{equation}
Our potential depends on scalar products alone. That means to determine the
torque we are required to know, \eg, that if
\begin{equation}
  \phi_1 = \frac12\Tr(\bq^\dagger \hat{z} \bq \hat{r}_{12})
\end{equation}
then
\begin{equation}
  \frac{\partial \phi_1}{\partial \bq}=\frac12 \bq^\dagger
  \left[\bq \hat{r}_{12} \bq^\dagger,\hat{z}\right]
\end{equation}
which we recognize as a cross product. We have taken, without loss of
generality, the reference vector to be $\hat{z}$ in the coordinate frame of the
particle. We then use the chain rule to differentiate $F_{12}$ with respect to
$\bq$, and convert back to Cartesian coordinates to find a contribution to the
torque of the form
\begin{equation}
  \vec{T}_{\phi_1}
  =-\hat r_{12}\times\hat n_1\frac{\partial F_{12}}{\partial \phi_1}
\end{equation}
Similar manipulations yield the other contributions.

\subsection{Numerical Methods}\label{Smethods-num}
\subsubsection{Monodisperse Systems} \label{sdmeth}\quad
For hard particle systems, the various contributions in Eq.~(5) (main text) can
be computed using different means. The pair interaction term ($\beta U$) is
given by the pair overlap function, which is known, at least in principle. The
pair Jacobian term ($-\log J$) can be computed analytically, as we describe
below. The third contribution ($\beta \tilde{F}_{12}$) could be computed by
determining the free energy of the rest of the system for a series of fixed
configurations of the pair. In practice, we are not interested in the
contributions of each of the individual terms \textit{per se}, so instead we
compute their sum directly. We obtain the PMFT on the left hand side of Eq.~(5)
(main text) by computing the frequency histogram of the relative pair
coordinates and orientations, and taking its logarithm, as suggested by the form
of Eq.~(4) (main text). This gives the PMFT up to an overall irrelevant additive
constant\footnote{Because the PMFT is a potential, only differences in the PMFT
between points are meaningful; it's value is not. Potential differences are not
affected by shifting the potential by a constant everywhere, so one can only
ever compute the potential up to a constant shift.}.

Since the PMFT is a generalization of the potential of mean force, the method
for computing the PMFT is a straightforward generalization of the method used to
extract the potential of mean force from the radial distribution function
$g(r)$. In detail, over the course of a MC simulation trajectory, we measure
all of the relative positions of each pair of particles that fall within a given
cutoff distance (for the results we present in the main text we
integrate over the relative orientations). We impose a discrete grid over the
set of allowed relative positions, and record the number of pairs that fall
within each grid cell. This tabulation of the relative frequency of the various
configurations, gives us a measurement of the $g(x,y,z)$ analogue of $g(r)$, and
so simply taking the logarithm gives us $F_{12}$ from Eq.\ (5) (main text).

Computing the PMFT in this manner introduces two forms of systematic
discretization error. To illustrate the sources of this error we will give a
more detailed description of the calculation we performed. We wish to compute
the PMFT at some given relative position and orientation. For concreteness, and
to keep formulae simple, let us consider just computing the force component. In
principle, one can compute the PMFT by performing thermodynamic integration or
umbrella sampling over various fixed relative particle positions if one wants
the exact potential difference between particular relative pair positions and
orientations. However, in our case we are interested in the general geometric
features of the potential, for which it is sufficient to perform simulations in
a standard ensemble, and record the relative frequency of various events. Thus
we compute an approximate value, $F'_{12}$, of the true PMFT, $F_{12}$, at some
point $(x_i,y_i,z_i)$ by averaging over a bin of size $(\Delta x,\Delta y,\Delta
z)$ centered at that point according to
\begin{equation}
  \Delta x\Delta y\Delta z e^{-\beta F'_{12}(x_i,y_i,z_i)}
  \equiv \int_{\text{bin}_i} dx dy dz e^{-\beta F_{12}(x,y,z)} \; .
\end{equation}
If the true potential of mean force and torque is slowly varying over the bin,
in the sense that
\begin{equation}
  \begin{split}
  \left|\int_{\text{bin}_i} dx dy dz (\vec x-\vec x_i) \cdot
  \nabla \left. e^{-\beta F_{12}(x,y,z)}\right|_{\vec x = \vec x_i}\right|\\
  \ll \Delta x\Delta y\Delta z e^{-\beta F_{12}(x_i,y_i,z_i)} \; ,
  \end{split}
\end{equation}
then we can Taylor expand the integrand about $(x_i,y_i,z_i)$, and, assuming
that the whole bin is allowed, we find that
\begin{equation}
  \Delta x\Delta y\Delta z e^{-\beta F'_{12}(x_i,y_i,z_i)}
  \approx \Delta x\Delta y\Delta z e^{-\beta F_{12}(x_i,y_i,z_i)} \; .
\end{equation}
This gives $F'_{12}(x_i,y_i,z_i)\approx F_{12}(x_i,y_i,z_i)$. This approximation
breaks down if $F_{12}$ is not sufficiently slowly varying, in the above sense.
It also breaks down if the bin is partially forbidden, \ie\ if the integrating
volume is not $\Delta x\Delta y\Delta z$. We expect the PMFT to vary quickly at
the boundaries of regions that are forbidden due to overlap, \ie\ we expect the
effective potential to be relatively ``hard''. This means that at the edges of
these regions, where we would have difficulty sampling events, we would also
expect to have to resolve minute differences to obtain meaningful results. This
makes such a technique prohibitively difficult for such boundary cases where one
would have to resort to umbrella sampling. As a result (in
particular in Figs.~2, 3, and 5, main text) we only show
the PMFT for points that are within $4\; k_\text{B}T$ of the global minimum,
because it is at these points that we can sample the potential reliably in the
sense described above, using our method.

\subsubsection{Penetrable Hard Sphere Systems}\label{depmeth}\quad
Here we explain how we calculate the PMFT for a pair of hard, arbitrarily shaped
colloidal particles in a sea of smaller penetrable hard sphere depletants.

As we showed in Eq.~(9) (main text), the PMFT takes on a simplified form in
the case of penetrable hard sphere depletants and thus the contribution from
integrating out the depletants $\tilde F_{12}$ can be computed more easily. The
contribution from the colloid pair potential is, in principle, known, as before,
and the Jacobian can be computed as described below. The
problem is therefore reduced to computing the contribution that comes from the
free volume available to the depletants for a fixed configuration of the
colloidal pair.

We computed the free volume in the penetrable hard sphere model of depletants by
MC integration. The same results could also be obtained by direct MC simulations
with explicit ideal depletants, as discussed below, or by
some other numerical integration method. To accelerate this computation, we make
note of the following. The change in free volume between a given configuration
and widely separated particles falls entirely within the intersection of two
spheres, each enclosing a particle. The volume of this region of intersection is
smallest (and, therefore, computation is most efficient) if the radii of the
spheres in question are as small as possible. This is given by having a sphere
enclosing each particle with a radius given by the sum of the radius of a sphere
that circumscribes the particle and the radius of a sphere that circumscribes
the depletant.

For a given relative position and orientation of the colloids, we computed the
intersection of spheres that determined this region and computed the change in
free volume by throwing random depletants uniformly over the region and
computing the fraction that intersected both colloids. We performed five
independent runs for each system. We used 500 MC integration points per unit
volume of the region of intersection, and 1000 total throws if the volume of
intersection was less than one unit. This fraction of the total volume gives the
change in free volume available to the depletants. In the case of faceted
spheres, overlap checks between particles were performed by casting the overlap
as an optimization problem that we solved using the Karush-Kuhn-Tucker method
\cite{KKT}. In the case of spherocylinders, the Bullet physics library
\cite{bullet} was used to check overlaps.

As noted above, the Jacobian of the transformation to invariant coordinates can
be computed analytically, which we did. We checked our analytical calculation
numerically by performing a MC simulation of a pair of free axisymmetric
particles. The frequency histogram of the occurrence of the invariant
coordinates was checked against the analytic form and found to match.

To examine these systems in detail it is convenient to use the language of
entropic patches from~\cite{epp}. We define the probability of specific binding
at a pair of entropic patch sites as the probability of finding a pair of
particles at the set of configurations that are in the basin of attraction of
perfect alignment. We obtain this by integrating the Boltzmann weight over the
set of configurations.

For concreteness, consider a pair of hemispherical particles in a bath of
depletants. The probability of specific binding can be cast formally as the
integral
\begin{equation}
  \begin{split}
    Z_s \propto & \int_{0}^{1} d\phi_1
    \int_{0}^{1} d\phi_2 \int dR d\chi
    e^{-\beta F_{12}(R,\phi_1,\phi_2,\chi)} H(-\beta P \Delta V_\text{F})
    \; .
  \end{split}
\end{equation}
The upper bounds on the $\phi$ integrals correspond to coincident faces. The
step $H$ function ensures that we are only integrating over configurations in
which there is entropic binding. Similar integrals can be defined for
semi-specific binding (patch to non-patch) and non-specific binding (non-patch
to non-patch).

\begin{figure}
  \begin{center}
    \input{tet}
  \end{center}
  \caption{An illustration of the choice of four sets of orthogonal coordinates
  for each of the faces of a tetrahedron, which we use for subsequent
  computations in Figs.~2 and 3 (main text). The face normals,
  which we take to be the \textcolor{gcb2}{$z$} coordinates in the frame of the
  face are shown in blue. Similarly, the \textcolor{gcb4}{$x$} and
  \textcolor{gcb6}{$y$} coordinates in the frame of the face are
  shown in green and red respectively.
  \label{tetcoord}}
\end{figure}
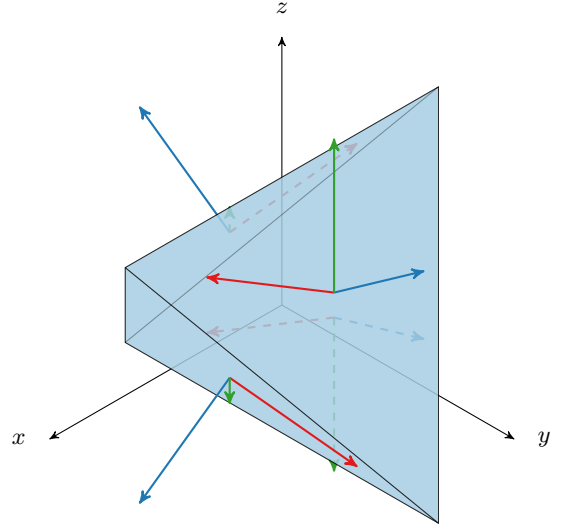
\begin{figure*}
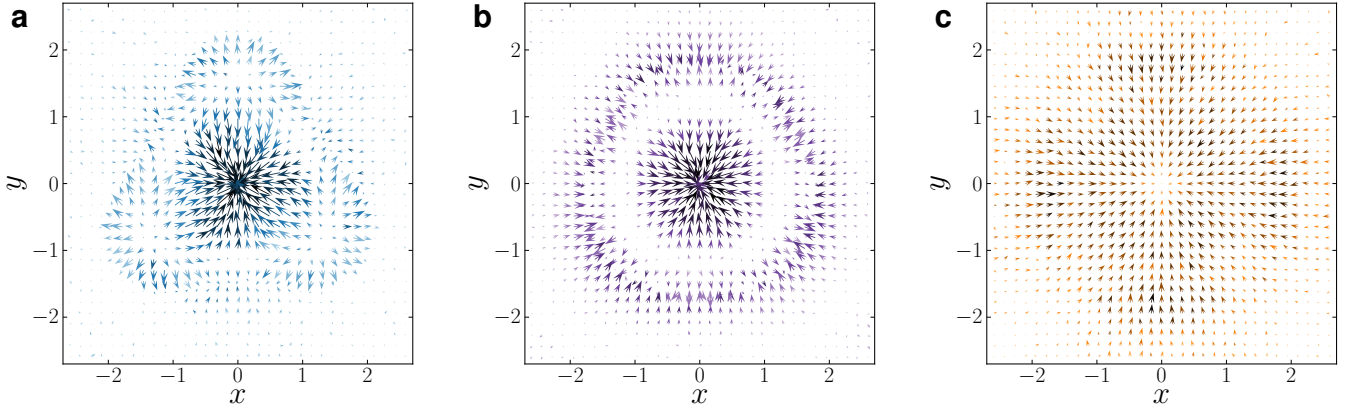

  \begin{tabular}{ccc}
    \includegraphics[width=6.0cm]{{{quiver_tet_0.4_2}}} &
    \includegraphics[width=6.0cm]{{{quiver_tf_0.4_2}}} &
    \includegraphics[width=6.0cm]{{{quiver_cube_0.4_3}}}
  \end{tabular}
  \caption{DEFs in Cartesian coordinates for monodisperse hard particle systems
  at a packing fraction of $\eta=0.4$ for tetrahedra (a), tetrahedrally facetted
  spheres (b), and cubes (c). DEFs are computed by approximating the negative
  gradient of the PMFT with finite differences. All cases show preference for
  face-to-face alignment, but as in Fig.\ 3 (main text) the strength of
  preference, and the shape of the forces, depends on particle shape.
  \label{fig-def}
  }
\end{figure*}
\subsubsection{Comparison of Methods for Penetrable Hard Sphere Depletants}\quad
\label{freevol}
The results we obtained via free volume calculations with ideal depletants can
also be obtained via simulations with explicit depletants. To see why the two
forms of computation are equivalent, we consider the following situation. Again,
for the sake of simplicity, we will work in the penetrable hard sphere limit.
The probability of accepting a trial Monte Carlo move of our colloidal particle
is given by
\begin{equation}
  p_a = (1-p)^N
\end{equation}
where $N$ is the number of depletants, and $p$ is the probability that a
depletant will be in the region swept out by the particle during its move. In
the limit in which we are working, this is given by
\begin{equation}
  p = \frac{\Delta V_\text{sweep}-\Delta V_\text{overlap}}{V_\text{F}}
\end{equation}
where $V_\text{F}$ is the free volume available to the depletants, $\Delta
V_\text{sweep}$ is the volume swept out by the colloid during move, and $\Delta
V_\text{overlap}$ accounts for any increase in the depletant overlap volume.

In the limit that the move is small, \ie\ $Np\ll 1$, the probability of
accepting the move is
\begin{equation}
  p_a \approx 1 - \frac{N(\Delta V_\text{sweep}-\Delta
  V_\text{overlap})}{V_\text{F}}
\end{equation}
which gives the probability of rejecting such a move as
\begin{equation}
  p_r \approx \frac{N(\Delta V_\text{sweep}-\Delta V_\text{overlap})}
  {V_\text{F}}
\end{equation}
We can, similarly, find the probability of rejecting a reverse move. That is
given by
\begin{equation}
  p_r' \approx \frac{N(\Delta V_\text{sweep})}
  {V_\text{F}+\Delta V_\text{overlap}}
  \approx \frac{N\Delta V_\text{sweep}}{V_\text{F}}
  \left(1-\frac{\Delta V_\text{overlap}}{V_\text{F}}\right)
\end{equation}
where we have assumed that, without loss of generality, the ``forward'' move
causes an increase in the depletant overlap volume, and the reverse move causes
it to decrease. We have again also used the fact that the size of the move is
small.

We compute the difference in probability for the two moves
\begin{equation}
  \Delta p \equiv p_r'-p_r \approx \frac{N\Delta V_\text{overlap}}
  {V_\text{F}}
  \left(1-\frac{\Delta V_\text{sweep}}{V_\text{F}}\right)
  \approx \frac{N\Delta V_\text{overlap}}{V_\text{F}} \; .
\end{equation}

We can now consider another pair of moves in which the initial
configuration of the forward move is identical to the situation just described,
but the final configuration is different. If in that case the change in overlap
volume is $\Delta V_\text{overlap}'$, then the probability difference is
\begin{equation}
  \Delta p' \approx \frac{N\Delta V_\text{overlap}'}{V_\text{F}}
\end{equation}

From these quantities we can compute the ratio $\Delta p/\Delta p'$, which is
dependent only on the free volume, which, in turn, is encoded in our potential
of mean force and torque. We get that
\begin{equation}
  \frac{\Delta p}{\Delta p'} =
  \frac{\Delta V_\text{overlap}}{\Delta V_\text{overlap}'}
\end{equation}
which means we can write
\begin{equation}
  \frac{\Delta p}{\Delta p'} = \frac{F_{12}^\text{post}-F_{12}^\text{pre}}
  {{F_{12}^\text{post}}'-F_{12}^\text{pre}}
\end{equation}
in the limit that $T\to0$. From this expression we see that the PMFT we deduced
from the free volume calculation is precisely the quantity that controls the
average acceptance rate of MC moves of the colloids in a simulation with
explicit depletants. Hence results obtained from the free volume methods used
above will precisely match those obtained using much more expensive MC
simulations with explicit ideal depletants.

\section{Supplementary Results}\label{Sresults}
\subsection{Entropic Forces In Monodisperse Hard Systems}
To capture DEFs, in the manuscript we computed the PMFT. Because the PMFT is a
potential, forces are the negative gradient of it. For completeness in Fig.\
\ref{fig-def}, we give explicit calculations of the DEFs for monodisperse hard
systems in Cartesian coordinates in three example systems at a packing fraction
of $\eta=0.4$: tetrahedra (a), tetrahedrally facetted spheres (b), and cubes (c).
The forces are computed using from the PMFT by approximating the gradient using
finite differences. We show the force components in the plane of the facet. To
produce the plots we have chosen to represent energies in units of length so
that arrows are visible on the plots. From Fig.\ 3 (main text) it can
immediately be seen that overall scale of forces is on the order of
$k_\mathrm{B}T/\sigma$ where $\sigma$ is the relevant length scale for the
particle. Panel (a) corresponds to Fig.\ 3c (main text), and shows that at
$\eta=0.4$, the vertices of the tetrahedra are repulsive, whereas the center of
the face is attractive. In contrast, panel (b) corresponds to Fig.\ 3g (main
text), and shows that removing the vertices of the tetrahedron has removed the
repulsion, but the center of the face is still attractive. Panel (c) corresponds
to Fig.\ 3k (main text) and shows (as does Fig.\ 3k) that the forces are less
strong for cubes than the other two particles at this density. Moreover, it
also shows that the cubic vertices are not acting repulsively at this density.

\subsection{Entropic Torque In Monodisperse Hard Systems}
In the main text we showed directional entropic forces between particles. Here
we give an explicit calculation of the torque that aligns particles. As a
simple example, consider a particle obtained from a sphere of radius $r$ by
cutting away the part of the sphere that intersects the half space
$\mathbb{R}^3$ for which $z/r>\alpha$ to We performed MC simulations of systems
of $1000$ such particles with $\alpha=0.01$ (nearly hemispherical) at fixed
volume. Since the particles have axial symmetry, we their relative position and
orientation can be characterized by four scalar quantities. If we take the
separation between the particles to be $\vec{q}_{12}$, and their symmetry axes
to be given by $\hat n_1$ and $\hat n_2$, then we are free to use
\begin{equation}
  \begin{split}
  \Delta \xi_{12} = \{ &
  R \equiv |\vec{q}_{12}|, \phi_1 \equiv \hat n_1 \cdot \hat q_{12},
  \phi_2 \equiv -\hat n_2 \cdot \hat q_{12},\\&
  \chi \equiv - \hat n_1 \cdot
  \hat n_2 \}
  \end{split}
\end{equation} 
In Table 
S1 we show the role of the PMFT in generating entropic
torques that align particle facets in this system at a density of $50\%$. We
find the potential difference between different particle orientations at fixed
separation distance is of the order of a few $k_\text{B}T$, giving rise to
entropic torques strongly favouring alignment. Figures are height of the
potential above the global minimum for slice of the potential with $\psi \equiv
\phi_1 = \phi_2 = \chi$ fixed. Although the angular differences are small
(perfect alignment is $\psi=1$), the effective interaction varies by more than
$2\; k_{B}T$ over this angular range at small separations, indicating that the
penalty for small misalignment is significant. Inset particle images illustrate
the relative orientations shown.

\begin{table*}
  \begin{tabular}{|c|cc|}
    \multicolumn{3}{l}{
    \small{\textbf{\textsf{Table S1. Angular dependence of PMFT for Hard
    Hemispheres}}}}
    \\
    \hline
    & $\psi\approx 0.95$ & $\psi\approx 0.98$ \\
    & \scalebox{0.05}{\includegraphics[angle=-5]{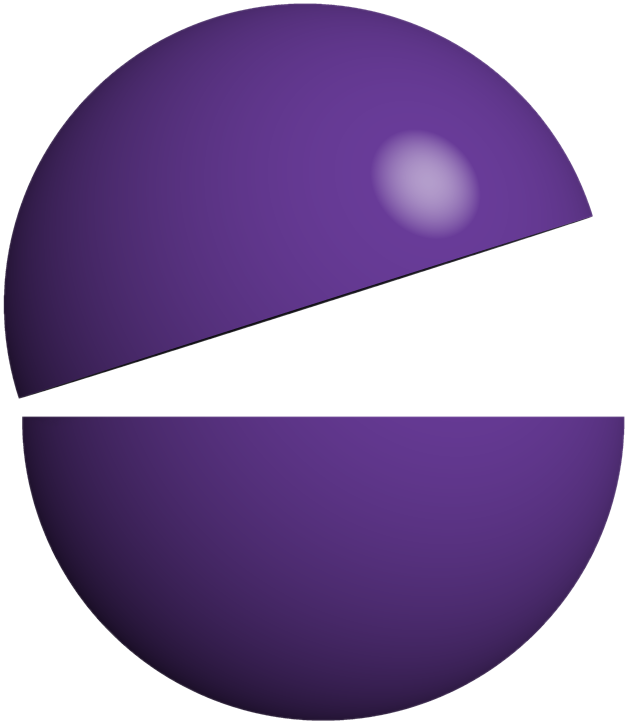}}
    & \scalebox{0.05}{\includegraphics[angle=-5]{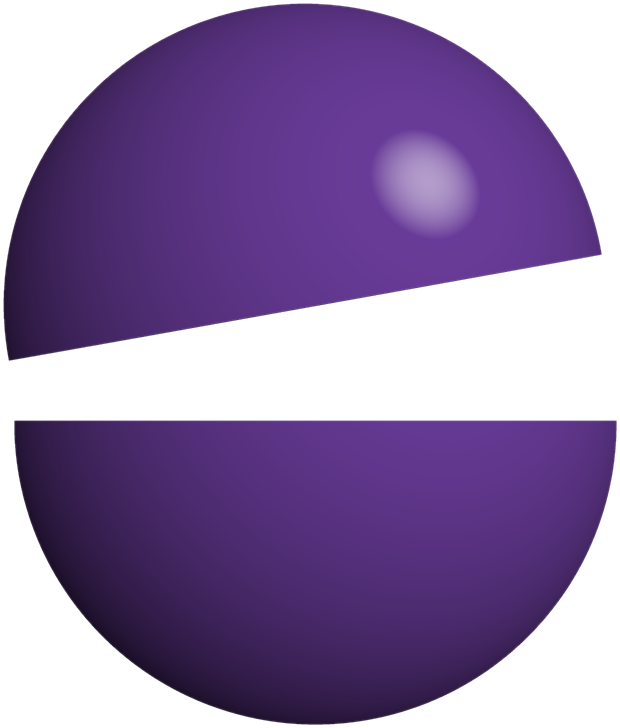}} \\
    \hline
    $F_{12}(R=0.3)/k_\mathrm{B}T$
    & $3.48 \pm 0.05$ & $1.16 \pm 0.01$ \\
    $F_{12}(R=0.4)/k_\mathrm{B}T$
    & $3.64 \pm 0.06$ & $2.39 \pm 0.03$ \\
    \hline
  \end{tabular}
\end{table*}
\subsection{Density Dependence}\label{density}
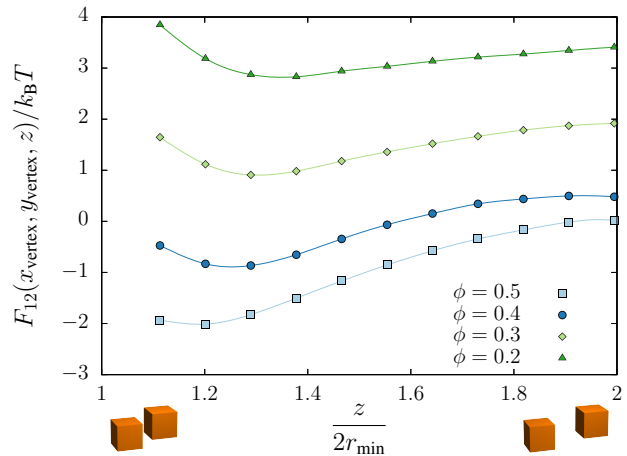
\begin{figure}
  \begin{center}
    \resizebox{8.6cm}{!}{\input{ze_cube}}
  \end{center}
  \caption{
  Density dependence of the PMFT along an axis perpendicular to the polyhedral
  face for a hard cube fluid that passes through the cube vertex. Note that the
  expected increase in effective attraction that occurs as system density
  increases is less pronounced than the corresponding plot in Fig.~2c (main
  text) for an axis that passes through the center of the face.
  \label{nozdep}
  }
\end{figure}

In Fig.~2 (main text) we showed that there was an effective attraction in
the direction perpendicular to the particle face that increased as the density
increased. We note here that this is not an effect of the increase in density
alone. For example, in Fig.~\ref{nozdep} we make a similar plot for cubes that,
rather than passing through the potential minimum, passes through the vertex of
one of the cubes. Comparing Fig.~\ref{nozdep} with Fig.~2c (main text), we
see that the effect of the particle shape leads to enhancement of face-to-face
contact over and above what we would expect to observe based solely on the
increase in density alone. This point is also underscored in Fig.~3 (main text)
above.

\section{Supplementary Discussion}\label{Sdiscuss}
\subsection{Entropy}
In this context, we should also comment further on the entropy that we are
computing when we compute the PMFT. We have used the fact that, at least in
principle, we can do statistical mechanics in any ensemble we find convenient.
As usual, the entropy of the system is computed by counting all the
microstates of the $N$ particles in the system, and this can be obtained by
performing (in the general case) the $6N$ dimensional integration over all of
the positions and orientations of all particles. Here, for the purposes of
isolating the effects of the particle shape, we have chosen to cast this
integral as a series of $6(N-1)$ dimensional integrals, each of which describes
the number of microstates available to the system when a particle pair is fixed
to a certain relative position and orientation \footnote{In practice, as
described below we approximate each of these contributions by
a $6N$ dimensional integral where six of the dimensions have a
near-infinitesimal domain of integration.}. It is in the comparison of the
relative entropic contributions from each of the integrals for pair
configurations that we are able to identify the ``microscopic'' entropic origin
of the ``macroscopic'' entropic ordering of the whole system seen in
\cite{epp}, and elsewhere in the literature. What is perhaps not intuitive
about this process in the hard particle limit, which is like a microcanonical
system in that all states have zero potential energy, is that this process
entails splitting up the ensemble into configurations of fixed particle pair
positions and orientations. We are, in effect, subdividing the microcanonical
ensemble into yet smaller isobaric ensembles of states. Computations of this
sort have appeared previously in the literature, see, \eg, \cite{svv,vijay}.

\subsection{Penetrable Hard Sphere Limit} \label{depl}
Because the sea particles have repulsive interactions with the pair of interest,
the pair behaves like part of the `box' that constrains the sea. By moving the
pair of interest we are changing the shape of the box from the point of view of
the sea particles. The force exerted by the sea particles on the pair, then, is
given by the stress tensor of the particles that are being integrated out on the
boundary defined by steric hindrance with the pair of interest. We note,
therefore, that the scale of this osmotic force is given by the scale of the
stress tensor in the system; if the stress tensor is isotropic then this is just
the pressure $P$, and so the scale of this force is given by
$P\sigma^3/k_\text{B}T$, where $\sigma$ is a characteristic length scale.

In fact, the DEFs in monodisperse systems, defined above, are further
strengthened by the addition of smaller soft depletants, as realized
experimentally in~\cite{youngmirkin,kayliedef}. Typical colloidal
experimental realizations of such systems require the depletants to induce
interactions with strength between $\sim4 \; k_\text{B}T$ (\eg\ \cite{mason})
and $\sim8 \; k_\text{B}T$ (\eg\ \cite{pacman}) to instigate binding. At very
high depletant concentrations, estimates of the effective interaction strength
can reach hundreds of $k_\text{B}T$ (\eg\ \cite{mason}). See
\cite{depletion} and references therein for more details on the experimental
measurement of depletion forces.

\subsection{Sea Particle Properties}
We have shown in general that hard systems have an entropic preference for more
densely packed pair configurations, because particle pairs are packed by the
thermal motion of sea particles. The existence of correlations among the sea
particles implies that the entropically preferred local dense packing for the
pair is not generally the same as the global densest packing for the pair.
However, if there are no correlations among the sea particles, their entropy
depends \emph{only} on the packing volume of the pair.

Intuitively, one can think of this as if the pair of interest is confined within
some membrane under external pressure, provided by the boundary of the sea
particles. If there is no correlation among the sea particles, the membrane will
behave as if it has no internal stiffness, and will deform to squeeze the pair
in any way they can be squeezed. However, if there are correlations among the
sea particles, the membrane will act as if it has some inherent structure that
prevents it from being deformed in any possible manner, and therefore the sea
particles may not be able to pack the pair into \emph{any} possible
configuration.

Also, if one considers an initially monodisperse hard system, and takes the
traditional (small) depletion limit at fixed system density, the number of sea
particles will increase dramatically. To leading order the contribution from the
sea particles scales like $-\beta N\log V$, which means that to preserve the
packing density of the system, as we scale the characteristic size of the
depletant $\sigma$, the sea particle contribution has scales (naively) as
$\sigma^{-3}$. This means that systems of colloids and traditional, small,
weakly interacting depletants, the osmotic pressure of the depletants can easily
be much greater than the osmotic pressure of the other colloids, in determining
the pair configuration. However, if the depletants are sufficiently large to be
completely excluded from the region within a colloidal aggregate, then the shape
entropy of the colloids within the aggregates then becomes important, as shown
in an experiment by Rossi \textit{et al.}\cite{rossi}.

\subsection{Many-Body Interactions}
The observation of face-to-face contacts in self-assembled systems, as noted
in~\cite{trunctet,zoopaper}, suggests that the effects of shape can be captured
by an effective pair potential, such as the PMFT. Indeed we have shown in this
paper a preference for polyhedra to align face-to-face, see Fig.\ 3 (main text).
However, there are reported systems where coincident face-to-face alignment is
not preferred, such as in octahedra (\eg\ \cite{trunctet}). We regard
these as many-body effects, which become more important at higher packing
fractions, and which would be captured only by a many-body PMFT.

Formally, the $n$-body extension of the present techniques is straightforward;
the only obstacle is to enumerate scalar invariant quantities for the $n$
bodies, and compute their Jacobian. In practice, however, as $n$ increases, so
does the difficulty of obtaining sufficiently many measurements to compute the
potential with accuracy.

\subsection{Interaction Range}
The range of DEFs is determined by the properties of the sea particles: their
intrinsic interactions, size, shape, \textit{etc.} Typically, we would expect
that the sea particles generate an effective interaction between the colloids
that is roughly of the order of the sea particle size. Given this intrinsic
limitation, one might ask whether we can design shape features on the colloidal
particles in order to control the assembly? This is developed in detail in
\cite{epp}; we briefly comment on it here, too.

If the sea particles are small, noninteracting depletants, the range is sufficiently
short that only shape features (on the colloidal pair) that are adjacent in
closely packed configurations contribute to the interaction, \eg\ flat
faces in polyhedra, dimples and complementary spheres in lock-and-key
experiments.\cite{lockkey,lockkeyent,pacman,pacsuprmol} Following \cite{epp},
we identify these features as ``entropic patches''. Note that if the patches are
sufficiently well-separated, the interactions are pair-wise additive, and so the
effective potential energy of the colloids is given by the sum of the pair
interaction energies.

If the sea particles are not small, or are interacting, then we would expect the
interaction range to be longer, and the approximation that individual particle
features can be considered separately as entropic patches may not always be
valid, though we have preliminary results in several cases suggesting it still
is.\cite{sphinx}

\end{document}

%% file: tet.tex
\begin{tikzpicture}[>=stealth']
  \coordinate (o) at (0,0);
  \coordinate (x) at ({1.7*cos(-150)},{1.7*sin(-150)});
  \coordinate (y) at ({1.7*cos(-30)},{1.7*sin(-30)});
  \coordinate (z) at (0,1.7);
  \coordinate (xc) at ($2.1*(x)$);
  \coordinate (yc) at ($2.1*(y)$);
  \coordinate (zc) at ($2.1*(z)$);
  \draw[->,draw=black] (o) -- (xc);
  \draw[->,draw=black] (o) -- (yc);
  \draw[->,draw=black] (o) -- (zc);
  \node[draw=none,fill=none,left=0.2cm of xc](xl){$x$};
  \node[draw=none,fill=none,right=0.2cm of yc](yl){$y$};
  \node[draw=none,fill=none,above=0.2cm of zc](zl){$z$};
  \coordinate (v1) at (${sqrt(2)}*(x)+(z)$);
  \coordinate (v2) at ($-{sqrt(2)}*(x)+(z)$);
  \coordinate (v3) at (${sqrt(2)}*(y)-1*(z)$);
  \coordinate (v4) at ($-{sqrt(2)}*(y)-1*(z)$);
  \coordinate (n1) at ($-{sqrt(2)}*(y)+(z)$);
  \coordinate (n2) at ($-{sqrt(2)}*(x)-1*(z)$);
  \coordinate (n3) at (${sqrt(2)}*(x)-1*(z)$);
  \coordinate (n4) at (${sqrt(2)}*(y)+(z)$);
  \coordinate (c1) at ($1.0/3.0*(n1)$);
  \coordinate (c2) at ($1.0/3.0*(n2)$);
  \coordinate (c3) at ($1.0/3.0*(n3)$);
  \coordinate (c4) at ($1.0/3.0*(n4)$);
  \coordinate (h1) at ($1.0/sqrt(3)*(n1)$);
  \coordinate (h2) at ($1.0/sqrt(3)*(n2)$);
  \coordinate (h3) at ($1.0/sqrt(3)*(n3)$);
  \coordinate (h4) at ($1.0/sqrt(3)*(n4)$);
  \coordinate (y1) at (${-sqrt(3)/2}*(x)+{0.5/sqrt(3)}*(y)+{1/sqrt(6)}*(z)$);
  \coordinate (y2) at (${-sqrt(3)/2}*(y)+{0.5/sqrt(3)}*(x)-{1/sqrt(6)}*(z)$);
  \coordinate (y3) at (${sqrt(3)/2}*(y)-{0.5/sqrt(3)}*(x)-{1/sqrt(6)}*(z)$);
  \coordinate (y4) at (${sqrt(3)/2}*(x)-{0.5/sqrt(3)}*(y)+{1/sqrt(6)}*(z)$);
  \coordinate (x1) at ($0.5*(x)+0.5*(y)+1.0/sqrt(2)*(z)$);
  \coordinate (x2) at ($0.5*(x)+0.5*(y)-1.0/sqrt(2)*(z)$);
  \coordinate (x3) at ($-0.5*(x)-0.5*(y)-1.0/sqrt(2)*(z)$);
  \coordinate (x4) at ($-0.5*(x)-0.5*(y)+1.0/sqrt(2)*(z)$);
  \draw[->,gcb2,thick] (c1) -- ($(c1)+(h1)$);
  \draw[->,gcb6,dashed,thick] (c1) -- ($(c1)+(y1)$);
  \draw[->,gcb4,dashed,thick] (c1) -- ($(c1)+(x1)$);
  \draw[fill=gcb1,opacity=0.6,draw=black,very thin,line join=round]
  (v1) -- (v2) -- (v4) --cycle;
  \draw[->,gcb2,dashed,thick] (c2) -- ($(c2)+(h2)$);
  \draw[->,gcb6,dashed,thick] (c2) -- ($(c2)+(y2)$);
  \draw[->,gcb4,dashed,thick] (c2) -- ($(c2)+(x2)$);
  \draw[fill=gcb1,opacity=0.6,draw=black,very thin,line join=round]
  (v2) -- (v3) -- (v4) --cycle;
  \draw[fill=gcb1,opacity=0.6,draw=black,very thin,line join=round]
  (v1) -- (v3) -- (v4) --cycle;
  \draw[->,gcb2,thick] (c3) -- ($(c3)+(h3)$);
  \draw[->,gcb6,thick] (c3) -- ($(c3)+(y3)$);
  \draw[->,gcb4,thick] (c3) -- ($(c3)+(x3)$);
  \draw[fill=gcb1,opacity=0.6,draw=black,very thin,line join=round]
  (v1) -- (v2) -- (v3) --cycle;
  \draw[->,gcb2,thick] (c4) -- ($(c4)+(h4)$);
  \draw[->,gcb6,thick] (c4) -- ($(c4)+(y4)$);
  \draw[->,gcb4,thick] (c4) -- ($(c4)+(x4)$);
\end{tikzpicture}

%% file: ze_cube.tex
\begingroup
  \makeatletter
  \providecommand\color[2][]{%
    \GenericError{(gnuplot) \space\space\space\@spaces}{%
      Package color not loaded in conjunction with
      terminal option `colourtext'%
    }{See the gnuplot documentation for explanation.%
    }{Either use 'blacktext' in gnuplot or load the package
      color.sty in LaTeX.}%
    \renewcommand\color[2][]{}%
  }%
  \providecommand\includegraphics[2][]{%
    \GenericError{(gnuplot) \space\space\space\@spaces}{%
      Package graphicx or graphics not loaded%
    }{See the gnuplot documentation for explanation.%
    }{The gnuplot epslatex terminal needs graphicx.sty or graphics.sty.}%
    \renewcommand\includegraphics[2][]{}%
  }%
  \providecommand\rotatebox[2]{#2}%
  \@ifundefined{ifGPcolor}{%
    \newif\ifGPcolor
    \GPcolortrue
  }{}%
  \@ifundefined{ifGPblacktext}{%
    \newif\ifGPblacktext
    \GPblacktexttrue
  }{}%
  \let\gplgaddtomacro\g@addto@macro
  \gdef\gplbacktext{}%
  \gdef\gplfronttext{}%
  \makeatother
  \ifGPblacktext
    \def\colorrgb#1{}%
    \def\colorgray#1{}%
  \else
    \ifGPcolor
      \def\colorrgb#1{\color[rgb]{#1}}%
      \def\colorgray#1{\color[gray]{#1}}%
      \expandafter\def\csname LTw\endcsname{\color{white}}%
      \expandafter\def\csname LTb\endcsname{\color{black}}%
      \expandafter\def\csname LTa\endcsname{\color{black}}%
      \expandafter\def\csname LT0\endcsname{\color[rgb]{1,0,0}}%
      \expandafter\def\csname LT1\endcsname{\color[rgb]{0,1,0}}%
      \expandafter\def\csname LT2\endcsname{\color[rgb]{0,0,1}}%
      \expandafter\def\csname LT3\endcsname{\color[rgb]{1,0,1}}%
      \expandafter\def\csname LT4\endcsname{\color[rgb]{0,1,1}}%
      \expandafter\def\csname LT5\endcsname{\color[rgb]{1,1,0}}%
      \expandafter\def\csname LT6\endcsname{\color[rgb]{0,0,0}}%
      \expandafter\def\csname LT7\endcsname{\color[rgb]{1,0.3,0}}%
      \expandafter\def\csname LT8\endcsname{\color[rgb]{0.5,0.5,0.5}}%
    \else
      \def\colorrgb#1{\color{black}}%
      \def\colorgray#1{\color[gray]{#1}}%
      \expandafter\def\csname LTw\endcsname{\color{white}}%
      \expandafter\def\csname LTb\endcsname{\color{black}}%
      \expandafter\def\csname LTa\endcsname{\color{black}}%
      \expandafter\def\csname LT0\endcsname{\color{black}}%
      \expandafter\def\csname LT1\endcsname{\color{black}}%
      \expandafter\def\csname LT2\endcsname{\color{black}}%
      \expandafter\def\csname LT3\endcsname{\color{black}}%
      \expandafter\def\csname LT4\endcsname{\color{black}}%
      \expandafter\def\csname LT5\endcsname{\color{black}}%
      \expandafter\def\csname LT6\endcsname{\color{black}}%
      \expandafter\def\csname LT7\endcsname{\color{black}}%
      \expandafter\def\csname LT8\endcsname{\color{black}}%
    \fi
  \fi
  \setlength{\unitlength}{0.0500bp}%
  \begin{picture}(7200.00,5040.00)%
    \gplgaddtomacro\gplbacktext{%
      \csname LTb\endcsname%
      \put(888,768){\makebox(0,0)[r]{\strut{}\large$-3$}}%
      \put(888,1337){\makebox(0,0)[r]{\strut{}\large$-2$}}%
      \put(888,1906){\makebox(0,0)[r]{\strut{}\large$-1$}}%
      \put(888,2475){\makebox(0,0)[r]{\strut{}\large$0$}}%
      \put(888,3044){\makebox(0,0)[r]{\strut{}\large$1$}}%
      \put(888,3613){\makebox(0,0)[r]{\strut{}\large$2$}}%
      \put(888,4182){\makebox(0,0)[r]{\strut{}\large$3$}}%
      \put(888,4751){\makebox(0,0)[r]{\strut{}\large$4$}}%
      \put(1032,528){\makebox(0,0){\strut{}\large$1$}}%
      \put(2179,528){\makebox(0,0){\strut{}\large$1.2$}}%
      \put(3326,528){\makebox(0,0){\strut{}\large$1.4$}}%
      \put(4473,528){\makebox(0,0){\strut{}\large$1.6$}}%
      \put(5620,528){\makebox(0,0){\strut{}\large$1.8$}}%
      \put(6767,528){\makebox(0,0){\strut{}\large$2$}}%
      \put(192,2759){\rotatebox{-270}{\makebox(0,0){\strut{}\Large$F_{12}(x_\text{vertex},y_\text{vertex},z)/k_\text{B}T$}}}%
      \put(3899,168){\makebox(0,0){\strut{}\Large$\dfrac{z}{2 r_\text{min}}$}}%
    }%
    \gplgaddtomacro\gplfronttext{%
      \csname LTb\endcsname%
      \put(5696,1671){\makebox(0,0)[r]{\strut{}\large$\phi=0.5$}}%
      \csname LTb\endcsname%
      \put(5696,1431){\makebox(0,0)[r]{\strut{}\large$\phi=0.4$}}%
      \csname LTb\endcsname%
      \put(5696,1191){\makebox(0,0)[r]{\strut{}\large$\phi=0.3$}}%
      \csname LTb\endcsname%
      \put(5696,951){\makebox(0,0)[r]{\strut{}\large$\phi=0.2$}}%
    }%
    \gplbacktext
    \put(0,0){\includegraphics{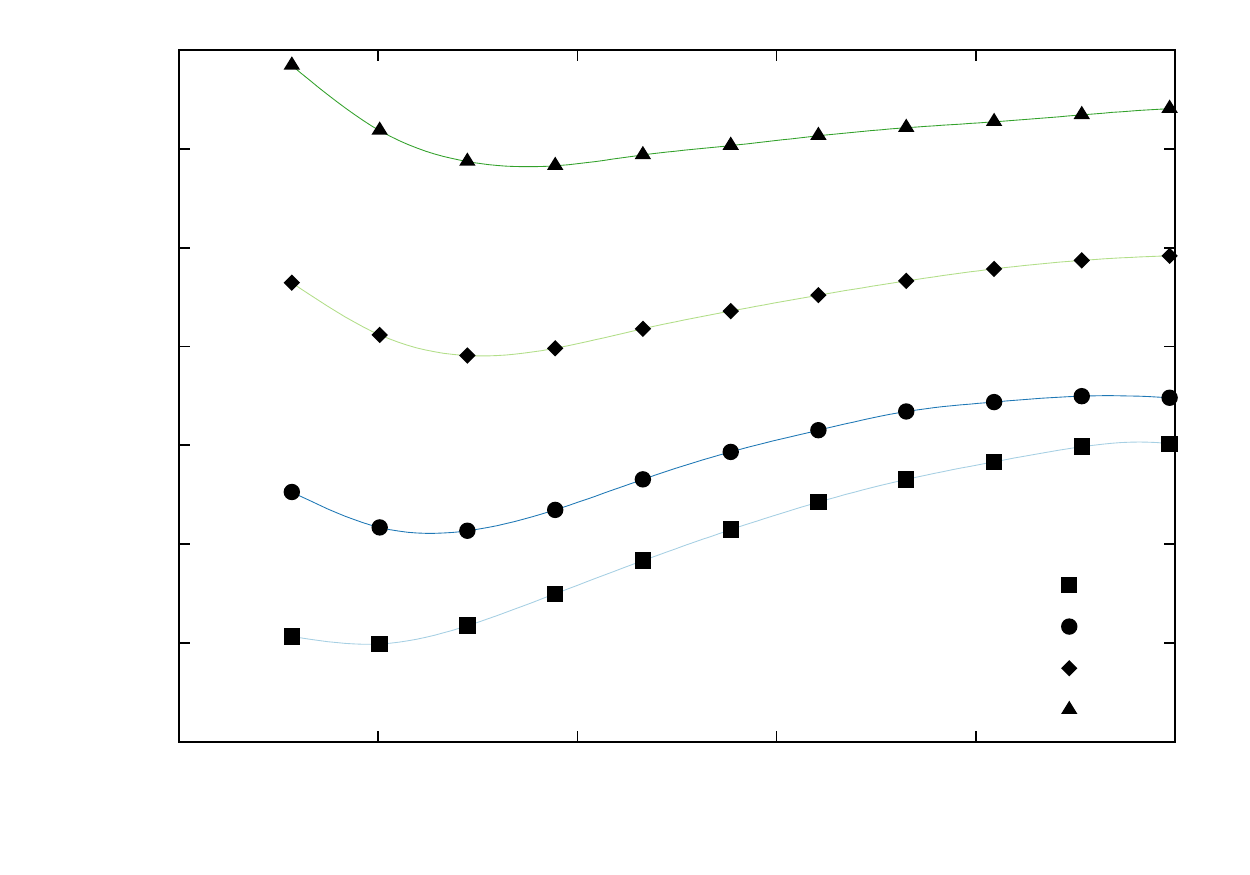}}%
    \put(0,0){\includegraphics{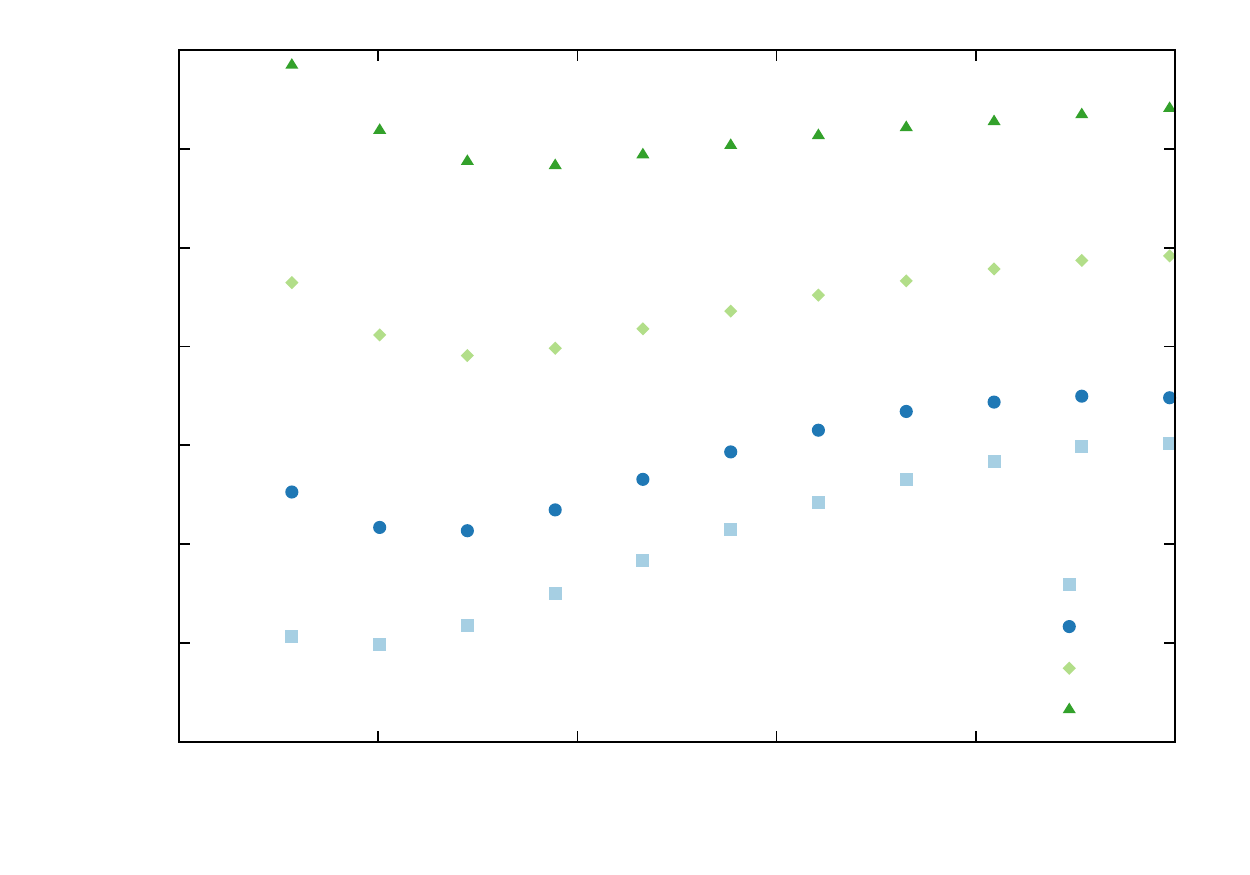}}%
    \put(1500,168){\makebox(0,0){\scalebox{0.05}{\includegraphics{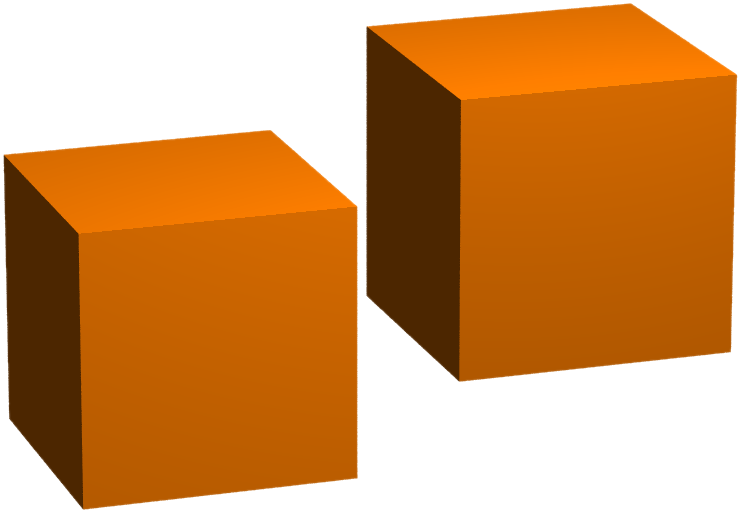}}}}
    \put(6200,168){\makebox(0,0){\scalebox{0.05}{\includegraphics{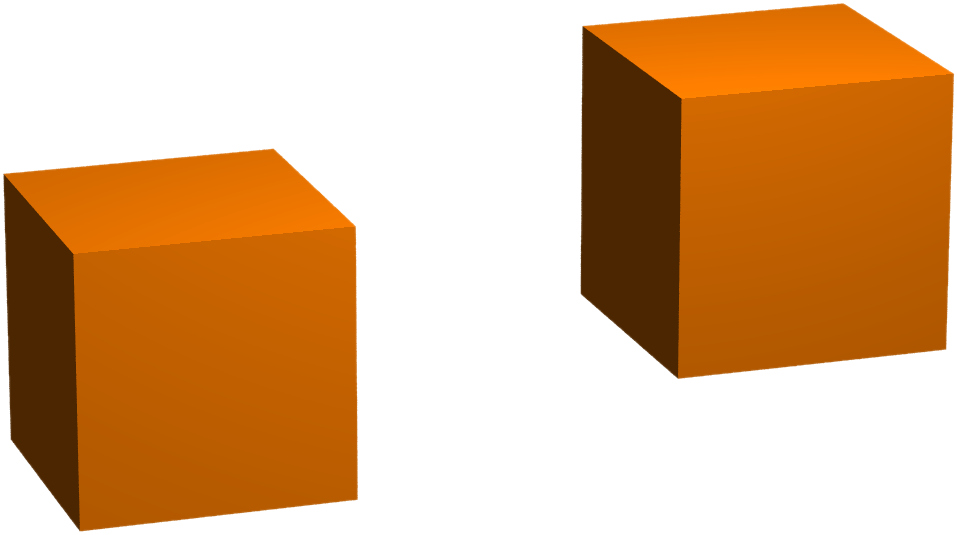}}}}
    \gplfronttext
  \end{picture}%
\endgroup